%% file: article.tex
\documentclass[floatfix,aps,twocolumn,showpacs,nofootinbib,superscriptaddress]{revtex4}

\usepackage{graphicx}
\usepackage{subfigure}
\usepackage{epsfig}
\usepackage{amsmath}
\usepackage{amsfonts}
\usepackage{amssymb}
\usepackage{bm}
\usepackage{hyperref}
\usepackage{dsfont}
\usepackage{lipsum}

\usepackage{tikz}
\usepackage{circuitikz}
\usetikzlibrary{arrows}

\newcommand{\mean}[1]{\left\langle #1 \right\rangle}
\newcommand{\meann}[1]{\langle #1 \rangle}

\newcommand{\beq}{\begin{equation}}
\newcommand{\eeq}{\end{equation}}

\DeclareMathOperator\Tr{Tr}

\newcommand\undermat[3][0pt]{%
  \makebox[0pt][l]{$\smash{\underbrace{\phantom{%
    \begin{matrix}\phantom{\rule{0pt}{#1}}#3\end{matrix}}}_{\text{#2}}}$}#3}

\begin{document}
\author{Nahuel Freitas}
\affiliation{Complex Systems and Statistical Mechanics, Physics and Materials Science,
University of Luxembourg, L-1511 Luxembourg, Luxembourg}
\author{Jean-Charles Delvenne}
\affiliation{Institute of Information and Communication Technologies, Electronics and
Applied Mathematics, Universit\'e catholique de Louvain, Louvain-La-Neuve, Belgium}
\author{Massimiliano Esposito}
\affiliation{Complex Systems and Statistical Mechanics, Physics and Materials Science,
University of Luxembourg, L-1511 Luxembourg, Luxembourg}


\title{Stochastic and Quantum Thermodynamics of Driven RLC Networks}

\date{\today}

\begin{abstract}
We develop a general stochastic thermodynamics of RLC electrical networks built on
top of a graph-theoretical representation of the dynamics commonly used by engineers.
The network is: {\it open}, as it contains resistors and current and voltage sources,
{\it nonisothermal}
as resistors may be at different temperatures, and {\it driven}, as circuit elements
may be subjected to external parametric driving.
The proper description of the heat dissipated in each resistor requires care within
the white noise idealization as it depends on the network topology.
Our theory provides the basis to design circuits-based thermal machines,
as we illustrate by designing a refrigerator using a simple driven circuit.
We also derive exact results for the low temperature regime in which the quantum
nature of the electrical noise must be taken into account. We do so using a
semiclassical approach which can be shown to coincide with a fully quantum
treatment of linear circuits for which canonical quantization is possible.
We use it to generalize the Landauer-B\"{u}ttiker formula for energy 
currents to arbitrary time-dependent driving protocols.
\end{abstract}

\maketitle

\section{Introduction}

Electronic circuits are versatile dynamical systems that can be designed
and built with a high degree of
precision in order to perform a great variety of tasks, from simple filtering and transmission
of analog signals to the complex processing of digital information in modern computers.
Traditional trends in the miniaturization of electronic components and in the increase of
operation frequencies are nowadays facing serious challenges related to the production of
heat and to the detrimental effects of thermal noise in the reliability of logical
operations \cite{kish2002}. Thus, thermodynamical considerations are central in
the search for new information processing technologies or improvements on the actual ones.
In light of this, it might be surprising that a general thermodynamical
description of electrical circuits is not available.


Perhaps one of the reasons for the absence of such general theory is the fact
that a satisfactory understanding of thermodynamics and fluctuations in
systems out of equilibrium has only been achieved in recent years. Stochastic
thermodynamics \cite{seifert2012,rao2018} is now emerging as a comprehensive framework
in which it is possible to describe and study thermodynamical processes
arbitrarily away from thermal equilibrium, and to obtain different `fluctuation
theorems' clarifying and constraining the role of fluctuations.
Simple electrical circuits have already been employed to experimentally
study non-equilibrium processes and to confirm the validity of fluctuation
theorems \cite{van2004, garnier2005, ciliberto2013, pekola2015}, however a general treatment is still lacking.

In this work we start by putting forward a general stochastic thermodynamic description of electrical
circuits composed of resistors, capacitors and inductors, as well as current and voltage sources.
The circuit components may parametrically depend on time due to an external controller.
Our theory makes use of the graph-theoretic description of the network dynamics in terms
of capacitor charges and inductor currents developed in electrical engineering \cite{balabanian1969, desoer2010}.
Noise is introduced via random voltage sources associated to resistors
which may lie at different temperatures.
For high temperatures, we use the standard Johnson-Nyquist noise which can be considered white.
The First and Second Law of thermodynamics are formulated based on an underdamped Fokker-Planck
description of the stochastic dynamics. A proper definition of local heat currents
(the rates at which energy is dissipated in each resistor) turns out to require some care.
The reason is that, depending on their topology, some circuits might display diverging
heat currents under the white noise idealization. This fact might be missed by
an analysis of the global energy budget alone, and is related to the anomalous
thermodynamic behaviour of overdamped models, that are however valid from a purely
dynamical point of view \cite{celani2012, polettini2013, bo2014, murashita2016}.
We clarify this issue and obtain a sufficient and necessary condition on the
topology of a circuit for it to be thermodynamically consistent.


After having established the high temperature theory, we proceed by demonstrating
how it can be used to design thermodynamic machines made of electrical circuits.
We do so by considering a simple electrical circuit where a resistor is cooled by
driving two capacitors connected to it in a periodic manner. Such schemes may be
employed to design new cooling strategies within electronic circuits.
This is particularly interesting in the quantum domain of low temperatures, where
parametrically driven circuits are commonly employed as low-noise amplifiers for the detection of small signals
down to the regime of single quantum excitations \cite{clerk2010,macklin2015}.
In fact, the possible use of these circuits as cooling devices has been pointed
out before \cite{niskanen2007,bergeal2010}. Thus, we generalize our theory to the low temperature quantum regime,
where the spectrum of the Johnson-Nyquist noise is not flat anymore
and is given by the Planck distribution. Our approach does not involve the
quantization of the degrees of freedom of the circuit, but can be shown to
be equivalent to an exact quantum treatment of linear circuits that have a
direct quantum analogue via canonical quantization \cite{schmid1982}. In this context we
derive a generalization of the Landauer formula for transport in non-driven
systems that is valid for arbitrary driving protocols. This is an important result
of this article as it provides an efficient computational tool that can be applied
to arbitrary circuits with any number of resistors at arbitrary temperatures, and
subjected to arbitrary driving protocols.
Finally, we show numerically that this formalism is able to capture
the quantum limits for cooling recently identified in \cite{freitas2017}.

%

This article is organized as follows. In Section \ref{sec:description_circuits} we
quickly review the basic graph-theoretical concepts involved in the description
of electrical circuits. This will serve also to introduce notation and to define
the basic objects to be employed later. In Section \ref{sec:deriv_dyn_eqs} we
derive the deterministic equation of motion for a given circuit and analyze
the energy balance and entropy production. In Section \ref{sec:stochastic_description}
we expand the deterministic description to consider the noise associated to
each resistive element. We provide an expression for the stochastic correction
to the local heat currents, which carry on to the energy balance and the
entropy production, and construct the Fokker-Planck equation describing the
stochastic evolution of the circuit state and heat currents.
In Section \ref{sec:example} we show how to apply our formalism
to study a simple example of an electrical heat pump.
Finally, in Section \ref{sec:quantum} we generalize our results to the low temperature
quantum regime.

\section{Description of RLC circuits}
\label{sec:description_circuits}

\begin{figure}
\includegraphics[scale=.17]{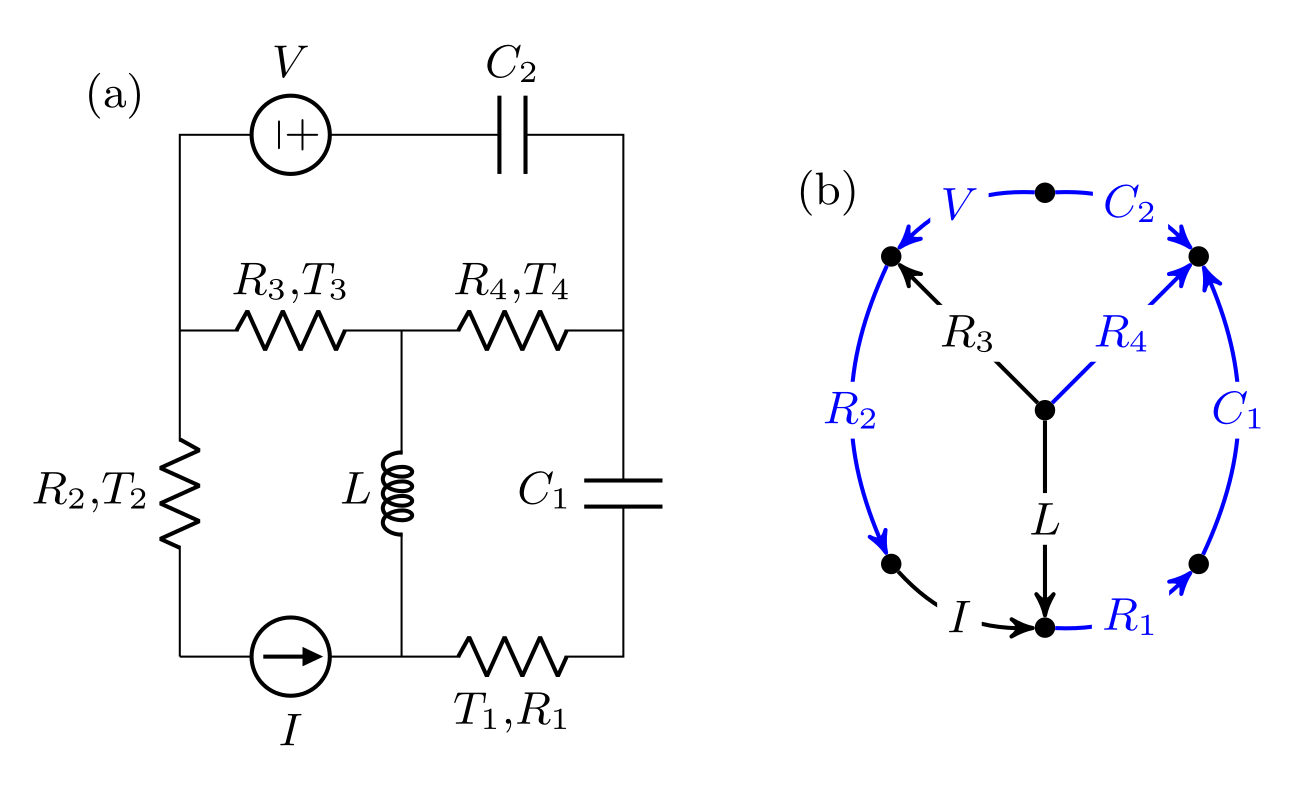}
\caption{(a) A circuit in which different resistors are at different temperatures.
(b) Its associated graph, with a normal tree indicated in blue.}
\label{fig:circuit}
\end{figure}

We consider circuits composed of two-terminal devices connected with each other
forming a network. A given circuit is mapped to a connected graph in which each two-terminal
device is represented by an oriented edge or branch (see Figure \ref{fig:circuit}).
The orientation of each edge
serves as a reference to indicate voltages drops and currents in the standard way.
The state of the circuit is specified by the nodes voltages
$u_1,\cdots,u_n$ and the edge currents $j_1, \cdots, j_b$,
where $n$ is the number of nodes and $b$ is the number of edges.
We consider five types of devices: voltage sources, current sources, capacitors, inductors
and resistors.
The charges of all capacitors and the magnetic fluxes of all inductors
are the dynamical variables of the circuit (alternatively, the voltages and
currents, respectively). In order for these variables to be truly independent,
we consider circuits fulfilling the following two conditions\footnote{
A given circuit can always be made to satisfy conditions (i) and (ii)
by adding small stray inductances or capacitances.}:
(i) The circuit graph has no loops\footnote{A loop or cycle is a sequence of edges
forming a path so that the first and last node coincide.}
formed entirely by capacitors and voltage sources,
and (ii) The circuit graph has no cut-sets\footnote{A cut-set or cocycle is a subset of edges such that their removal splits
the graph into at least two disconnected parts. As a loop, it can be oriented, with the orientation
indicating a prefered direction from one of the disconnected parts to the other.}
formed entirely by inductors and current sources.
If condition (i) is fulfilled, then the voltages of all capacitors and voltage sources
are independent variables, i.e., they cannot be directly related via the
Kirchhoff's voltage law (KVL).
Analogously, if condition (ii) is fulfilled, then the currents of all inductors
and current sources are independent variables since they are not directly constrained
via the Kirchhoff's current law (KCL).

The mathematical description of the circuit is easily constructed based on a \emph{tree}
of the circuit graph. A tree is a fully connected subgraph with no loops (see Figure \ref{fig:circuit}-(b)).
Under conditions (i) and (ii) it is always possible to find a tree for which all
the edges corresponding to capacitors and voltage sources are part of the tree,
and all the edges corresponding to inductors and current sources are out of it.
We will call such
a tree a \emph{normal tree} and we will base our dynamical description of
the circuit on it.
Following the terminology of \cite{balabanian1969} we will
refer to edges in the normal tree as \emph{twigs} and edges outside it as \emph{links}
(also known as co-chords and chords, respectively). Thus, all capacitors and
voltage sources are twigs, while all inductors and current sources are links.
The edges corresponding to resistors can be split in two groups
according to whether or not they are part of the normal tree.
Every tree has $b_t = n-1$ twigs and therefore $b_l = b-(n-1)$ links.

\subsection{Loops and cut-sets matrices}

Given a set of oriented loops we define the loop matrix $B$ as follows:
\begin{equation}
B_{i,j} = \begin{cases}
1 \text{ if edge $j$ is in loop $i$ with $=$ orientation}\\
-1 \text{ if edge $j$ is in loop $i$ with $\neq$  orientation }\\
0 \text{ otherwise}
\end{cases}
\end{equation}
A normal tree can be used to define a set of independent oriented loops in the
following way: take the tree subgraph and add a link edge to it, then a loop will
be formed (otherwise that link should have been a twig). The orientation of the
loop is chosen to coincide with that of the added link. This can be done for
every link. If we order the edges by counting first the twigs and then the links,
the loop matrix thus obtained has the following structure:
\begin{equation}
B = \left[
\begin{array}{cc}
B_\text{twig} & \mathds{1}_{b_l} \\
\end{array}
\right],
\label{eq:matrix_B}
\end{equation}
where $\mathds{1}_{k}$ is the $k\times k$ identity matrix.
For the normal tree indicated in Figure \ref{fig:circuit}-(b), the matrix $B_\text{twig}$ is:
\begin{equation}
B_\text{twig} =
\begin{array}{c|cccccc}
 & \textcolor{blue}{V} & \textcolor{blue}{C_1} & \textcolor{blue}{C_2} & \textcolor{blue}{R_1} & \textcolor{blue}{R_2} & \textcolor{blue}{R_4}\\
 \hline
R_3 & -1 & 0 & 1 & 0 & 0 & -1\\
L  & 0 & 1 & 0 & 1 & 0 & -1\\
I  & 1 & 1 & -1 & 1 & 1 & 0
\end{array},
\end{equation}
and specifies which twigs are involved in the loops corresponding to each link.

Given a set of oriented cut-sets (or cocycles) we define the cut-set matrix $Q$ as:
\begin{equation}
Q_{i,j} =
\begin{cases}
1 \text{ if edge $j$ is in cut-set $i$ with $=$ orientation}\\
-1 \text{ if edge $j$ is in cut-set $i$ with $\neq$  orientation }\\
0 \text{ otherwise}
\end{cases}
\end{equation}
As before, a normal tree can be used to define a set of independent cut-sets.
The procedure is as follows. Take the tree subgraph and remove a twig, then
the graph is split into two disconnected subgraphs. Consider all the edges
going from one subgraph to the other, including the removed twig. These edges
then form a cut-set which is oriented as the removed twig. This is then repeated
for every twig, obtaining the following cut-set matrix:
\begin{equation}
Q = \left[
\begin{array}{cc}
\mathds{1}_{b_t} & Q_\text{link}\\
\end{array}
\right].
\label{eq:matrix_Q}
\end{equation}
With the same ordering as before, the matrix $Q_\text{link}$ for the circuit of
Figure \ref{fig:circuit} is:
\begin{equation}
Q_\text{link}   =
\begin{array}{c|ccc}
& R_3 & L & I \\
\hline
\textcolor{blue}{V} &  1 & 0 & -1\\
\textcolor{blue}{C_1} &  0 & -1 & -1\\
\textcolor{blue}{C_2} &  -1 & 0 & 1\\
\textcolor{blue}{R_1} &  0 & -1 & -1\\
\textcolor{blue}{R_2} &  0 & 0 & -1\\
\textcolor{blue}{R_4} &  1 & 1 & 0
\end{array},
\label{eq:qlink_example}
\end{equation}
and as we see specifies which links belong to the cut-set corresponding to a
given twig. More details about the construction of the loop and cutset matrices
can be found in Appendix \ref{ap:description_circuits}.

The matrices $B$ and $Q$ are orthogonal in the sense that $BQ^T = 0$. This
property follows naturally from their definitions and implies that $B_\text{twig} = -Q_\text{link}^T$.
We also mention that loops and cut-sets can be identified algebraically from the incidence
matrix of the graph, and that the loop and cut-set matrices satisfy additional
algebraic relations \cite{polettini2015}.

\subsection{Kirchhoff's Laws and Tellegen's theorem}

If $j$ and $v$ are column vectors with the edge current and voltage \emph{drops} as components,
respectively, then the Kirchhoff's laws can be expressed in terms of the loop
and cut-set matrices as follows:
\begin{subequations}
\begin{align}
\text{KVL:}\qquad B v &= 0\\
\text{KCL:}\qquad Q j &= 0.
\end{align}
\label{eq:kirchoff}
\end{subequations}
These form a set of $b_t + b_l = b$ independent algebraic equations that can be used
to eliminate half of the $2b$ variables contained in $j$ and $v$. In particular,
if we order the components of $j$ and $v$ such that twig variables appear first,
we can employ Eqs. (\ref{eq:matrix_B}) and (\ref{eq:matrix_Q}) and write:
\begin{subequations}
\begin{align}
\qquad v_l &= -B_\text{twig}\: v_t = Q_\text{link}^T\: v_t\\
\qquad j_t &= -Q_\text{link}\: j_l,
\end{align}
\end{subequations}
where $v^T = [v_t^T, v_l^T]$ and $j^T = [j_t^T, j_l^T]$. We see then that it is
enough to give the twigs voltages and the links currents to determine the rest
of the variables.

From the orthogonality of $B$ and $Q$ and the two Kirchhoff's laws, it is possible
to prove Tellegen's theorem: any vector $v$ compatible with the KVL and
any vector $j$ compatible with the KCL are orthogonal, i.e, $j^Tv = 0$.

\subsection{Block structure of $j$, $v$ and $Q_t$}
In the following we will assume the previous splitting of the vectors
$j$ and $v$:
\begin{equation}
j=\left[
\begin{array}{c}
j_t\\
j_l
\end{array}
\right]
\qquad
v=\left[
\begin{array}{c}
v_t\\
v_l
\end{array}
\right].
\end{equation}
Also, since all voltage sources (E) and capacitors (C) are twigs, while all current
sources (I) and inductors (L) are links, we can further split the vectors $j$ and $v$
as follows:
\begin{equation}
x_t=\left[
\begin{array}{c}
x_E\\
x_C\\
x_{R_t}
\end{array}
\right]
\qquad
x_l=\left[
\begin{array}{c}
x_{R_l}\\
x_L\\
x_I
\end{array}
\right],
\end{equation}
where $x$ stands for $j$ or $v$ and $R_t$ and $R_l$ indicate the resistive edges
which are respectively twigs or links.
This partitioning induces the following block structure in the matrix $Q_\text{link}$:
\begin{equation}
Q_\text{link} = \left[
\begin{array}{ccc}
Q_\text{ER} & Q_\text{EL} & Q_\text{EI} \\
Q_\text{CR} & Q_\text{CL} & Q_\text{CI} \\
Q_\text{RR} & Q_\text{RL} & Q_\text{RI} \\
\end{array}
\right].
\label{eq:block_Q}
\end{equation}
Thus, for example, each column of the block $Q_\text{CL}$ correspond to an inductive edge (which is a link)
and its rows specify which capacitive edges (twigs) are involved in the cut-set corresponding to that inductive edge.
For the matrix of Eq. \eqref{eq:qlink_example}, we have $Q_\text{CL} = \left[ \begin{smallmatrix} -1 \\ 0 \end{smallmatrix}\right]$.

\subsection{Constitutive relations}

Each two-terminal device in the circuit is characterized by a particular
relation between the electric potential difference across its terminals and the current
through it. Voltage sources just fix a definite value for the potential difference,
regardless of the current, and current sources fix a current value regardless of the
voltage. Resistors are described by an algebraic relation between voltage and current.
We can write,
\begin{equation}
v_{R_t} = R_t \: j_{R_t} \qquad v_{R_l} = R_l \: j_{R_l},
\label{eq:const_R}
\end{equation}
where $R_t$ and $R_l$ are diagonal matrices with the resistances of the twigs
and links resistors as non-zero elements, respectively. $R_t$ and $R_l$ can be
time-dependent. Also, for non-linear resistors like diodes these matrices can
also be functions of the voltages or currents.

Finally, capacitors and inductors are described by the following set of differential
equations:
\begin{equation}
\frac{d}{dt}(C\:v_C) = j_C \qquad \frac{d}{dt}(L\: j_L) = v_L.
\end{equation}
Here, $C$ and $L$ are symmetric matrices describing the capacitances and inductances
of the circuit. $C$ is usually diagonal, but $L$ could account for cross-couplings
between different inductors. As with resistors, they can also depend on time
and/or describe non-linearities.

\section{Derivation of dynamical equations}
\label{sec:deriv_dyn_eqs}

We want to obtain an equation of motion for the circuit describing the evolution
of the voltages of all capacitors and the currents in all inductors. We begin
with the constitutive equations for them:
\begin{equation}
\frac{d}{dt}
\left[
\begin{array}{c}
C\: v_C\\
L\: j_L
\end{array}
\right]
=
\left[
\begin{array}{c}
j_C\\
v_L
\end{array}
\right].
\end{equation}
The task now is to use the Kirchhoff's laws and the algebraic constitutive
equations for the resistors to express the variables $j_C$ and $v_L$ in terms
of the dynamical ones $v_C$ and $j_L$. First, we use the following KCL and KVL
equations which are parts of Eq. \eqref{eq:kirchoff}:
\begin{subequations}
\begin{align}
-j_C &= Q_\text{CR} \: j_{R_l} + Q_\text{CL}\: j_L + Q_\text{CI} \:j_I\\
v_L &= Q_\text{EL}^T \: v_{E} + Q_\text{CL}^T\: v_C + Q_\text{RL}^T \:v_{R_t},
\end{align}
\label{eq:deriv_kirchoff_1}
\end{subequations}
and we then obtain:
\begin{equation}
\begin{split}
&
\frac{d}{dt}
\left[
\begin{array}{c}
C\: v_C\\
L\: j_L
\end{array}
\right]
=
\overbrace{
\left[
\begin{array}{cc}
& -Q_\text{CL}\\
Q_\text{CL}^T &
\end{array}
\right]
}^{\mathcal{M}_c}
\left[
\begin{array}{c}
v_C\\
j_L
\end{array}
\right]+\\
&
\underbrace{
\left[
\begin{array}{cc}
&-Q_\text{CI}\\
Q_\text{EL}^T &
\end{array}
\right]
}_{\mathcal{M}_s}
\left[
\begin{array}{c}
v_{E}\\
j_{I}
\end{array}
\right]
+
\underbrace{
\left[
\begin{array}{cc}
-Q_\text{CR}&\\
& Q_\text{RL}^T
\end{array}
\right]
}_{\mathcal{M}_d^T}
\left[
\begin{array}{c}
j_{R_l}\\
v_{R_t}
\end{array}
\right],
\end{split}
\label{eq:def_Q_1}
\end{equation}
where we have defined the matrices $\mathcal{M}_c$, $\mathcal{M}_s$ and
$\mathcal{M}_d$ (the subindices stand for \emph{conservative}, \emph{sources}
and \emph{dissipation}, as will be justified in the following).
We now only need to eliminate $j_{R_l}$ and $v_{R_t}$, since $v_E$ and $j_I$
are given. For this we use the constitutive relations of Eq. (\ref{eq:const_R})
and two additional Kirchhoff relationships:
\begin{subequations}
\begin{align}
-R_t^{-1} \: v_{R_t} &= Q_\text{RR} \: j_{R_l} - Q_\text{RL} \: j_L - Q_\text{RI} \: j_I\\
R_l \: j_{R_l} &= Q_\text{ER}^T \: v_{E} - Q_\text{CR}^T \: v_C - Q_\text{RR}^T \: v_{R_t},
\end{align}
\label{eq:deriv_kirchoff_2}
\end{subequations}
which can be rewritten as:
\begin{equation}
\begin{split}
\overbrace{
\left[
\begin{array}{cc}
R_l & -Q_\text{RR}^T\\
Q_\text{RR} & R_t^{-1}
\end{array}
\right]
}^{{\alpha^{-1}}}
\left[
\begin{array}{c}
j_{R_l}\\
v_{R_t}
\end{array}
\right]
&=
\overbrace{
\left[
\begin{array}{cc}
Q_\text{ER}^T &\\
& -Q_\text{RI}
\end{array}
\right]
}^{-\mathcal{M}_{sd}}
\left[
\begin{array}{c}
v_E\\
j_I
\end{array}
\right]\\
&+
\left[
\begin{array}{cc}
Q_\text{CR}^T&\\
&-Q_\text{RL}
\end{array}
\right]
\left[
\begin{array}{c}
v_{C}\\
j_{L}
\end{array}
\right],
\end{split}
\label{eq:def_Q_2}
\end{equation}
where we have defined the additional matrices $\alpha$ and $\mathcal{M}_{sd}$
(in this case the subindex stands for \emph{source dissipation}).
Inserting this relation in Eq. (\ref{eq:def_Q_1}), a closed dynamical
equation for the variables $v_C$ and $j_L$ is obtained. We can express
it in the following concise form:
\begin{equation}
\frac{d}{dt}
\left[
\begin{array}{c}
C\: v_C\\
L\: j_L
\end{array}
\right]
=
\mathcal{A}(t)
\left[
\begin{array}{c}
v_{C}\\
j_{L}
\end{array}
\right]
+
\mathcal{B}(t)
\left[
\begin{array}{c}
v_{E}\\
j_{I}
\end{array}
\right],
\end{equation}
where the matrix coefficients are given by:
\begin{equation}
\mathcal{A}(t) = \mathcal{M}_{c} - \mathcal{M}_{d}^T
{\alpha}(t)\mathcal{M}_{d},
\label{eq:def_A}
\end{equation}
and
\begin{equation}
\mathcal{B}(t) = \mathcal{M}_{s} - \mathcal{M}_{d}^T
{\alpha}(t)\mathcal{M}_{sd}.
\label{eq:def_B}
\end{equation}
Note that ${\alpha}$ might depend on time if resistances do.
Some comments about the structure of the matrix $\mathcal{A}$ are in order.
The first term in Eq. (\ref{eq:def_A}), $\mathcal{M}_{c}$,
describes the conservative interchange of
energy between capacitors and inductors. In some cases it can be interpreted as analogous
to the symplectic matrix in Hamiltonian mechanics. It has a block structure that stems
from the separation of variables according to its behaviour under time reversal
(voltages are even under time reversal while currents are odd).
The second term takes into account the effect of the resistances in the dynamics.
It is twofold: the antisymmetric part of
$\mathcal{M}_{d}^T {\alpha}(t)\mathcal{M}_{d}$ describes
the interchange of energy between capacitors and inductors that is allowed by
resistive channels, and the symmetric part describes the loss of energy of those
elements (see Eq. \ref{eq:def_E_diss}). Importantly, both the symmetric and
antisymmetric part of $\mathcal{M}_{d}^T {\alpha}(t)\mathcal{M}_{d}$
respect the block structure of $\mathcal{M}_{c}$. The meaning of this
property is that energy dissipation is bound to be an invariant quantity upon time
reversal, and has important consequences regarding the non-equilibrium
thermodynamic behaviour of the circuit, as discussed in more detail in
Appendix \ref{ap:block_struct}.

\subsection{Charge and flux variables}

Instead of working with $v_C$ and $j_L$ as dynamical variables, from
a physical point of view it is more natural to work with the charges in the
capacitors and the magnetic fluxes in the inductors. They are defined as $q = C\: v_C$ and
$\phi = L \: j_L$, respectively. We will group them in a column vector
$x$. Thus, we can write
\begin{equation}
x =
\left[
\begin{array}{c}
q\\
\phi
\end{array}
\right]
=
\underbrace{
\left[
\begin{array}{cc}
C&\\
&L
\end{array}
\right]
}_{\mathcal{H}^{-1}}
\left[
\begin{array}{c}
v_C\\
j_L
\end{array}
\right],
\end{equation}
where we have defined the matrix $\mathcal{H}$.
With these definitions, the dynamical state equation now reads:
\begin{equation}
\frac{dx}{dt}
=
\mathcal{A}(t) \mathcal{H}(t)\: x
+
\mathcal{B}(t) s(t),
\end{equation}
where $s^T = [v_E^T, j_I^T]$ is a vector grouping the voltage and currents of the
sources, that can depend on time.

\subsection{Linear energy storage elements}
If we consider the particular case in which the matrices $C$ and $L$,
and thus also $\mathcal{H}$, do not depend on the state vector $x$, we
can express the energy contained in the circuit at a given time as
a quadratic function of the circuit state:
\begin{equation}
E(x,t) = \frac{1}{2} \; x^T \: \mathcal{H}(t) \: x.
\label{eq:linear_energy}
\end{equation}
Then, we can write the dynamical state equation as:
\begin{equation}
\frac{dx}{dt}
=
\mathcal{A}(t) \: \nabla E(x,t)
+
\mathcal{B}(t) s(t).
\label{eq:dyn_linear_storage}
\end{equation}
Note that Eq. (\ref{eq:dyn_linear_storage}) still allows for non-linear
resistive relations, and in that case the matrices $\mathcal{A}$ and $\mathcal{B}$
have a tacit dependence on $x$ (through ${\alpha}$).

\subsection{Energy balance}

We now analyze the balance of energy between the different elements
of the circuit and the entropy produced during its operation. We consider
linear storage elements since in this case we have a simple notion of energy
associated to a given state $x$ of the circuit, which is given by Eq. (\ref{eq:linear_energy}).
We begin by writing down the variation in time of the circuit energy:
\begin{equation}
\begin{split}
\frac{d}{dt}E(x,t) &= \nabla E(x,t)^T \frac{dx}{dt} + \frac{\partial }{\partial t} E(x,t) \\
&=
\underbrace{
\nabla E^T \mathcal{A}(t) \nabla E
}_\text{dissipation}
+
\underbrace{
\nabla E^T \mathcal{B}(t) s(t)
}_\text{forcing}
+
\underbrace{
\frac{\partial E}{\partial t}
}_\text{driving}.
\end{split}
\end{equation}
We see that it naturally splits into tree distinct terms that account for
different mechanisms via which the energy stored in capacitors and inductors
can change. The first one describes how this energy is dissipated into the
resistors. Note that only the symmetric part of $\mathcal{A}$ plays a role
in the expression $\nabla E^T  \mathcal{A} \: \nabla E$, and from the definition
of $\mathcal{A}$ in Eq. (\ref{eq:def_A}), we see that it can only be different
from zero if there are resistors in the circuit. Explicitly,
\begin{equation}
\dot E_\text{diss} =  \nabla E^T  \mathcal{A} \: \nabla E  =
- \nabla E^T  \mathcal{M}_{d}^T \: ({\alpha})_s  \: \mathcal{M}_{d} \nabla E,
\label{eq:def_E_diss}
\end{equation}
where $(X)_s$ indicates the symmetric part of $X$.
Secondly, voltage and current
sources in the circuit can give energy to the capacitors and inductors, and
this is described by the second term. Finally, external changes in the capacitances
or inductances can also contribute to the circuit energy. This is clearly
identified as work performed on the circuit by an external agent.

To complete the understanding of the energy balance of the circuit, we analyze
the total rate of energy dissipation in the resistors (i.e, Joule heating).
Thus, the instantaneous dissipated power in each resistor is $\dot Q_r = j_r v_r$,
and the total rate of heat production reads:
\begin{equation}
\dot Q  =
\sum_r \dot Q_r =
\left[
\begin{array}{cc}
j_{R_l}^T & v_{R_t}^T
\end{array}
\right]
\overbrace{
\left[
\begin{array}{cc}
R_l & \\
& R_t^{-1}
\end{array}
\right]
}^{R}
\left[
\begin{array}{c}
j_{R_l} \\ v_{R_t}
\end{array}
\right] \geq 0 .
\label{eq:def_matrix_R}
\end{equation}
Using Eq. (\ref{eq:def_Q_2}) to eliminate the resistor currents and voltages we
find:
\begin{equation}
\begin{split}
\dot Q &= \nabla E^T \mathcal{M}_{d}^T \: {\alpha}^T R{\alpha} \: \mathcal{M}_{d} \nabla E \\
&+(\mathcal{M}_{sd} \:s+ 2\mathcal{M}_{d} \nabla E)^T
\: {\alpha}^T R{\alpha} \: \mathcal{M}_{sd} \:s.
\end{split}
\label{eq:diss_heat}
\end{equation}
As we will see next, the first term of the previous expression is exactly $-\dot E_\text{diss}$
defined in Eq. \eqref{eq:def_E_diss}, and thus represents the part of the energy dissipated
into the resistors that is lost by capacitors and inductors. The second term
corresponds to the part of the energy that is dissipated into the resistors directly by the
voltage or current sources. 
The identity between $-\dot E_\text{diss}$ and the first term of Eq. (\ref{eq:diss_heat})
can be established from the following property of the matrix ${\alpha}$:
\begin{equation}
{\alpha}^T R {\alpha} = {\alpha} R {\alpha}^T = ({\alpha})_s,
\label{eq:symm_alpha}
\end{equation}
that can be proven from the definition of ${\alpha}$
via $2\times 2$ block matrix inversion. Using this, we can write the following
expression for the energy balance of the circuit:
\begin{equation}
\frac{dE}{dt} = -\dot Q + \dot W = -\dot Q + \dot W_s + \dot W_d.
\label{eq:energy_balance}
\end{equation}
where the total work rate $\dot W$ is the sum of
the rates of work performed by the sources, $\dot W_s$, and
by an external agent that drives the circuit parameters, $\dot W_d$.
They are given by:
\begin{equation}
\dot W_d = \frac{\partial E}{\partial t} = \frac{1}{2} x^T \frac{d\mathcal{H}}{dt} x,
\end{equation}
and
\begin{equation}
\begin{split}
\dot W_s &= \nabla E^T \mathcal{B}(t) s(t) \\&+
(\mathcal{M}_{sd} \:s+ 2\mathcal{M}_{d} \nabla E)^T
\: {\alpha}^T \! R{\alpha} \: \mathcal{M}_{sd} \:s.
\end{split}
\label{eq:work_sources}
\end{equation}
As expected, using the fact that
$j_C^T v_C + j_L^T v_L = \frac{dE}{dt} - \frac{\partial E}{\partial t}$
and Tellegen's theorem, we find that
\begin{equation}
j_E^T v_E + j_I^T v_I = -\dot W_s.
\end{equation}

\subsection{Entropy production}

At the level of description considered so far, the only aspect of entropy
production that can be accounted for is the one related to the generation of
heat in the resistors due to non-vanishing net currents. Thus, to every resistor $r$
we associate a instantaneous entropy production equal to $\dot \Sigma_r =\dot Q_r/T_r$,
where $T_r$ is the temperature of the considered resistor.
Therefore, we can write the total entropy production as
\begin{equation}
\dot \Sigma = \sum_r \frac{j_r v_r}{T_r} =
\left[
\begin{array}{cc}
j_{R_l}^T & v_{R_t}^T
\end{array}
\right]
R\:
\beta
\left[
\begin{array}{c}
j_{R_l} \\ v_{R_t}
\end{array}
\right] \geq 0 ,
\label{eq:noiseless_entropy}
\end{equation}
where $\beta$ is a diagonal matrix with the inverse temperatures of the resistors as non-zero
elements.
As with $\dot Q$, Eq. (\ref{eq:def_Q_2}) can be employed to express $\dot \Sigma$
in terms of known quantities.
However, we know in advance that the previous expression for entropy production
is incomplete, since it doesn't take into account the effects of fluctuations.
There are three main ways or mechanisms via which the fluctuations can play a role.
In general, if there are fluctuations then the state of the circuit and its evolution
become stochastic. Therefore an entropy can be associated to the probability distribution
of the state $x$ at any time, which will in general change and evolve in a non-trivial way.
Of course, if the circuit is operating in a steady state this internal contribution to the
total entropy production will vanish. But even in that situation fluctuations can
still play a role, in combination with other two conditions. First, if the
resistors are at different temperatures then fluctuations alone can transport
heat from a hot resistor to a cold one \cite{ciliberto2013, ciliberto2013b}, and this non-equilibrium
process has an associated entropy production. Secondly, even if all
the resistors are at the same temperature, heat transport can be induced by the
driving of the circuit parameters \cite{delvenne2014}. For example, it is possible to devise
a cooling cycle in which two capacitors are used as a working medium
to extract heat from a resistor and dump it into another resistor (see Section
\ref{sec:example}).
In linear systems, this can be done only through fluctuations, without affecting the
mean values of current and voltages and even in the case in which they vanish
at all times. All these aspects of the full entropy production in linear circuits
will be explored using stochastic thermodynamics in the next sections.

\section{Stochastic dynamics}
\label{sec:stochastic_description}

\subsection{Johnson-Nyquist noise}
In this section we extend the previous description of RLC circuits to
take into account the effects of the Johnson-Nyquist noise \cite{johnson1928,nyquist1928}
originating in the resistors. To do so, we model a real resistor as a noiseless
resistor connected in series with a random voltage source. The voltage
$\Delta v$ of this source has zero mean, $\mean{\Delta v(t)} = 0$, and for high
temperatures is modeled as delta correlated:
\begin{equation}
\mean{\Delta v(t)\Delta v(t')} = 2R k_bT \delta(t-t'),
\end{equation}
where $R$ is the resistance of the considered resistor, $T$ is its temperature,
and $k_b$ is the Boltzmann constant. This corresponds to a flat noise spectrum
$S(\omega) = R k_bT/\pi$. For low temperatures or high frequencies,
one should instead consider the `symmetric' \cite{schmid1982, clerk2010} quantum noise
spectrum:
\begin{equation}
S(\omega) = \frac{R}{\pi} \frac{\hbar \omega}{2} \coth\left(\frac{\hbar \omega}{2 k_b T}\right)
= \frac{R}{\pi} \hbar\omega \left(N(\omega) + 1/2\right),
\label{eq:noise_spectrum}
\end{equation}
where $N(\omega)=(\exp(\hbar\omega/(k_bT))-1)^{-1}$ is the Planck's distribution.
The 1/2 term added to the Planck's distribution takes into account the environmental
ground state fluctuations. It has no consequences regarding the thermodynamics
of static circuits, even for non-equilibrium conditions. However, it does have
important consequences
in the ultralow temperature regime in the case of driven circuits, as discussed
in detail in Section \ref{sec:periodic_driving}. In the following we will just
consider the classical case of high temperatures and delta correlated noise,
where simpler results hold and the usual machinery of stochastic calculus can
be employed.
Then we present in Section \ref{sec:quantum} the full treatment of
quantum noise using Green's functions techniques.

\subsection{Langevin dynamics}

Thus, if we want to consider Johnson-Nyquist noise in our general description of
circuits we need to introduce a random voltage source with the previous characteristics
for every resistor in the given circuit. This can be done directly by considering
each voltage source as a new element in the circuit graph, or indirectly by
taking into account the presence of the new voltage sources in the Kirchhoff's
laws of the original circuit. The latter approach has the advantage of not modifying
the graph of the circuit and is the one we are going to use in the following.
Thus, Eq. (\ref{eq:deriv_kirchoff_1}-b) becomes
\begin{equation}
v_L = Q_\text{EL}^T \: v_{E} + Q_\text{CL}^T\: v_C + Q_\text{RL}^T \: (v_{R_t}+ \Delta v_{R_t}),
\end{equation}
and similar modifications to Eqs. (\ref{eq:deriv_kirchoff_2})
lead to the following generalization of Eq. (\ref{eq:def_Q_2}):
\begin{equation}
\begin{split}
{\alpha}^{-1}\!\!
\left[
\begin{array}{c}
j_{R_l}\\
v_{R_t}
\end{array}
\right]\!
=
-\mathcal{M}_{sd} \: s -\mathcal{M}_{d} \mathcal{H} \: x
\!+\!
\left[
\!\!
\begin{array}{c}
-\Delta v_{R_l} \!+\! Q_\text{RR}^T \Delta v_{R_t}\\
0
\end{array}
\!\!
\right]\!,
\end{split}
\label{eq:resistor_variables_stoch}
\end{equation}
where $\Delta v_{R_l}$ and $\Delta v_{R_t}$ are column vectors with
the random voltage associated with links or twigs resistors, respectively.
These modifications to the Kirchhoff's laws are propagated straightforwardly to
the final equation of motion in Eq. (\ref{eq:dyn_linear_storage}), which now reads:
\begin{equation}
\frac{dx}{dt}
=
\mathcal{A}(t) \: \nabla E(x,t)
+
\mathcal{B}(t) \: s(t)
+ \mathcal{M}_{d}^T \; {\alpha}(t) \; \eta(t),
\label{eq:sto_linear_storage}
\end{equation}
where the vector $\eta(t)$ groups the random voltages in the following way:
\begin{equation}
\eta(t) =
\left[
\begin{array}{c}
-\Delta v_{R_l} \\
R_t^{-1} \: \Delta v_{R_t}
\end{array}
\right].
\end{equation}
In what follows, for simplicity, we will omit the explicit dependence
on time of $\mathcal{A}$, $\mathcal{B}$ and $\alpha$,
which are anyway constant if the resistances are also constant.
The previous expression for the equation of motion can be cast into a
slightly more symmetric form. For this we note that since the statistics of the
noise variables are invariant under sign inversion we can neglect the minus sign in front of
$\Delta v_{R_l}$. Also, for high temperatures, taking into account the form
of the spectrum for each random voltage we can write:
\begin{equation}
\eta(t) = \sqrt{2 R \: k_b \beta^{-1} } \: \xi(t),
\label{eq:def_eta}
\end{equation}
where $R$ is the matrix defined in Eq. (\ref{eq:def_matrix_R}), $\beta$ is
the matrix of inverse temperatures defined in Eq. (\ref{eq:noiseless_entropy}),
and $\xi(t)$ is a column vector of unit-variance white noise variables (as many
as there are resistors).
Then, the equation of motion for the circuit takes the following Langevin form:
\begin{equation}
\frac{dx}{dt} =
\mathcal{A} \: \nabla E(x,t)
+
\mathcal{B} \: s(t)
+ \sum_r \sqrt{2k_b T_r} \: \mathcal{C}_r \:\xi(t),
\label{eq:langevin}
\end{equation}
where $\mathcal{A}$ and $\mathcal{B}$ are the same as in the deterministic case.
The last sum is over every resistor
in the circuit, and $\mathcal{C}_r$ is given by
\begin{equation}
\mathcal{C}_r = \mathcal{M}_{d}^T {\alpha}  R^{1/2} \Pi_r,
\label{eq:def_Cr}
\end{equation}
where $\Pi_r$ is a projector over the one-dimensional subspace corresponding to
resistor $r$. Note that the matrices $\mathcal{C}_r$
depend on time only if resistances do. Also, they always satisfy
$\mathcal{C}_r \mathcal{C}_{r'}^T = 0$ for $r\neq r'$. However the other
possible product $\mathcal{C}_{r}^T\mathcal{C}_{r'}$ does not vanish in general.
Eq. (\ref{eq:langevin}) is the stochastic generalization of the usual state
equation for electrical circuits.

%

\subsection{Mean values and covariance matrix dynamics}
\label{sec:evol_cov_matrix}
For linear circuits, the mean values $\mean{x}$ of the charges and fluxes
evolve according to the deterministic equation
of motion:
\begin{equation}
\frac{d\mean{x}}{dt} = \mathcal{A} \mathcal{H}(t) \mean{x} + \mathcal{B} \: s(t).
\label{eq:mean_values_evol}
\end{equation}
The evolution of the covariance matrix $\sigma(t) = \mean{(x-\mean{x})(x-\mean{x})^T}$ can be easily derived
from the Langevin equation and reads:
\begin{equation}
\frac{d}{dt} \sigma(t) = \mathcal{A}\mathcal{H}(t) \sigma(t) +
\sigma(t) \mathcal{H}(t)\mathcal{A}^T + \sum_r 2k_bT_r \: \mathcal{C}_r\mathcal{C}_r^T.
\label{eq:cov_matrix_evol}
\end{equation}
It is important to note that for linear circuits the evolution of the covariance matrix
is completely independent from the forcing function $s(t)$ (if it is noiseless, as we considered).
Thus, deterministic forcing of the circuit via voltage or current sources can
only change the mean values of the charges and fluxes in the circuit, but not
the fluctuations around them nor their correlations.

\section{Stochastic thermodynamics}

\subsection{Fluctuation-dissipation relation}

We now consider the particular case in which the circuit is not driven
(its parameters are time independent), all the resistors are at the same
temperature $T$, and no voltages or currents are applied ($s=0$). Then,
according to equilibrium thermodynamics the system must attain a thermal state for
long times, with a distribution:
\begin{equation}
p_\text{th} (x) \propto  e^{-\frac{1}{2k_bT} x^T \mathcal{H} x}.
\end{equation}
This is a Gaussian state with covariance matrix
\begin{equation}
\sigma_\text{th} = k_b T \: \mathcal{H}^{-1}.
\end{equation}
It is instructive to check that in fact this result is obtained from the
previous dynamical description of the circuit. Thus, if there is an asymptotic
stationary state we see from Eq. (\ref{eq:cov_matrix_evol}) that in isothermal
conditions its covariance matrix $\sigma_\text{st}$ must satisfy
\begin{equation}
0 = \mathcal{A}\mathcal{H} \sigma_\text{st} +
\sigma_\text{st} \mathcal{H}\mathcal{A}^T +
2 k_b T \sum_r \: \mathcal{C}_r\mathcal{C}_r^T.
\label{eq:cov_matrix_st}
\end{equation}
Also, using the definition of the matrices $\mathcal{C}_r$ (Eq. \eqref{eq:def_Cr}),
we see that:
\begin{equation}
\begin{split}
\sum_r \: \mathcal{C}_r\mathcal{C}_r^T
&=
\mathcal{M}_{d}^T {\alpha} R {\alpha}^T \mathcal{M}_{d}
=
\mathcal{M}_{d}^T (\alpha)_s \mathcal{M}_{d}\\
&= - \frac{\mathcal{A}+\mathcal{A^T}}{2} = - (\mathcal{A})_s,
\end{split}
\label{eq:fd_relation}
\end{equation}
where in the second equality we used the previously mentioned property
${\alpha} R {\alpha}^T = ({\alpha})_s$. Eq. \eqref{eq:fd_relation} is nothing
else that the fluctuation-dissipation (FD) relation for general RLC circuits.
Using the FD relation, we can easily verify that the covariance matrix
$\sigma_\text{th}$ is in fact a solution of Eq. (\ref{eq:cov_matrix_st}).

We finally note that, due to the linearity of the circuit, if the circuit is forced
by constant voltages and/or current sources, then the asymptotic state is given
by a displaced thermal state:
\begin{equation}
p_\text{disp} (x) \propto  e^{-\frac{1}{2k_bT} (x-\mean{x})^T \mathcal{H} (x-\mean{x})},
\label{eq:displaced_thermal}
\end{equation}
where the mean values $\mean{x}$ are given by the stationary solution of Eq. \eqref{eq:mean_values_evol}.
However, $p_\text{disp}$ is not an equilibrium state since,
at variance with $p_\text{th}$, it has a non-vanishing entropy production
given by Eq. \eqref{eq:noiseless_entropy}.

\subsection{Energy balance}

To write down the energy balance for the stochastic description, we begin
by noticing that since the energy is a quadratic function, its mean value
can be easily expressed in terms of the mean values $\mean{x}$ and the covariance matrix $\sigma$:
$\mean{E(x,t)} = E(\mean{x},t) + \Tr[\mathcal{H}\sigma]/2$. Therefore, we have
\begin{equation}
\frac{d}{dt}\mean{E} = \frac{d}{dt}E(\mean{x},t) +
\frac{1}{2}\Tr\left[\frac{d\mathcal{H}}{dt}\sigma\right]+
\frac{1}{2}\Tr\left[\mathcal{H}\frac{d\sigma}{dt}\right],
\end{equation}
where $E(x,t)$ is the energy function in Eq. \eqref{eq:linear_energy}.
Thus, the energy balance takes the same form as before
\begin{equation}
\frac{d}{dt}\mean{E}=  -\dot{\mean{Q}}+ \dot{\mean{W_s}} + \dot{\mean{W_d}},
\label{eq:energy_balance_2}
\end{equation}
with the following expressions for the driving work and heat terms:
\begin{equation}
\dot{\mean{W_d}} = \frac{\partial}{\partial t} E(\mean{x},t) + \frac{1}{2} \Tr\left[ \frac{d\mathcal{H}}{dt} \sigma(t)\right],
\end{equation}
and
\begin{equation}
\dot{\mean{Q}} = \dot Q (\mean{x}, t) - \frac{1}{2}\Tr\left[\mathcal{H}\frac{d\sigma}{dt}\right],
\label{eq:stoch_Q}
\end{equation}
where $\dot Q(x,t)$ is the heat rate of Eq. \eqref{eq:diss_heat}.
Finally, the work performed by the sources is equal to its deterministic value
in Eq. \eqref{eq:work_sources}
(for noiseless sources):
\begin{equation}
\dot{\mean{W_s}} = \dot{W}_s(\mean{x},t).
\end{equation}

\subsection{Local heat currents}

Using Eq. (\ref{eq:cov_matrix_evol}) for $d\sigma/dt$ and the FD relation
of Eq. (\ref{eq:fd_relation}), we can rewrite
Eq. (\ref{eq:stoch_Q}) for the total heat dissipation rate as:
\begin{equation}
\meann{\dot Q} = \sum_r \left( \mean{j_r} \! \mean{v_r}
+ \Tr[(\mathcal{H} \sigma \mathcal{H} - k_b T_r \mathcal{H}) \mathcal{C}_r\mathcal{C}_r^T]
\right),
\end{equation}
where for clarity we omitted the explicit time dependence of $\mathcal{H}$.
We see that each term in the sum of the previous
equation can be associated to a particular resistor in the circuit. Then, they
are a sensible definition for the local heat currents $\meann{\dot Q_r}$
(the rate of energy dissipation in resistor $r$):
\begin{equation}
\meann{\dot Q_r} =\mean{j_r} \! \mean{v_r}
+ \Tr[(\mathcal{H} \sigma \mathcal{H} - k_b T_r \mathcal{H}) \mathcal{C}_r\mathcal{C}_r^T].
\label{eq:local_heat}
\end{equation}
This heuristic definition of the local heat currents is sometimes considered in
the literature (see, for example, \cite{parrondo1996}), but it is not always correct.
In fact, if we add to the quantities $\meann{\dot Q_r}$ defined in the previous
equation a term of the form $\sum_{r'} \Delta \dot Q_{r,r'}$, for any antisymmetric
tensor $\Delta \dot Q_{r,r'}$, we find that the total heat rate $\meann{\dot Q}=\sum_r
\meann{\dot Q_r}$ remains unchanged. Then, it is in general not possible to derive
local heat currents via a decomposition of the global one.

\begin{figure}
\includegraphics[scale=.19]{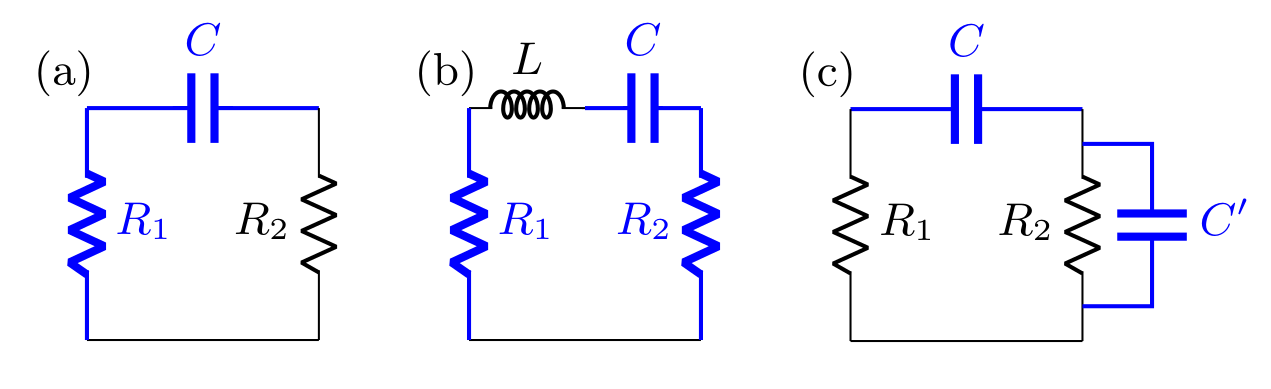}
\caption{(a) A simple circuit with diverging local heat currents in the white
noise limit, due to the possibility of high frequency fluctuations originating in one resistor to
be dissipated in the other one. (b-c) Two possible modifications, where high
frequency fluctuations are filtered out by a series inductor (b) or a parallel capacitor (c).
The edges of a normal tree in each case are shown in blue. Note that Eq. (\ref{eq:local_heat})
could be applied to the circuit in (a) to obtain well behaved quantities, that
however do not represent the actual heat currents. In fact, applying the same
equation to the circuit in (b) we obtain the correct heat currents $\meann{\dot Q_{1/2}}
=\pm k_b \Delta T R/L$ (for $R_1 = R_2 = R$ and $\Delta T = |T_1 - T_2|$),
that diverge in the limit $L\to 0$. Note that the topological condition $Q_\text{RR}=0$
is able to distinguish the circuits in (b) and (c) from the circuit in (a).}
\label{fig:topology_condition}
\end{figure}

From another perspective, in the context of electrical circuits the local
heat currents can be naturally defined as
\begin{equation}
\meann{\dot Q_r} = \mean{j_r (v_r + \Delta v_r)},
\label{eq:def_local_heat}
\end{equation}
where $\Delta v_r$ is the random voltage associated to each resistor. That quantity,
however, is found to be divergent in the general case.
The physical origin for this divergence is the fact that for some
circuits thermal fluctuations of arbitrarily high frequencies originating in a resistor
(which are always present in the model due to the white-noise idealization) can
be dissipated into another, as illustrated in Figure \ref{fig:topology_condition}
with a simple example. There are two different reasons why this might happen.
On one hand, the initial description of the circuit might be
missing important degrees of freedom that are relevant for the
thermodynamics, while still being valid from a purely dynamically point of view.
This was recently discussed in \cite{celani2012, polettini2013, bo2014,
murashita2016}, in connection to the overdamped approximation in stochastic dynamics,
even though not in the specific context of electrical circuits.
In particular, in \cite{murashita2016} it was shown that when a particle is
in simultaneous contact with multiple reservoirs at different temperatures there
is a transfer of heat associated to its momentum degree of freedom
at a rate that is inversely proportional to the relaxation
timescale, which is taken to be infinitesimal in the overdamped limit
(Figure \ref{fig:topology_condition}-(b) is actually an example of this situation,
as explained in the caption).
On the other hand, heat currents might also be ill-defined independently of whether
or not some kind of underdamped approximation was made in the description of
the circuit. In those cases one should consider a more realistic model for the
noise in the resistors, going beyond the white noise idealization. This is the more
general way to address this issue, which is discussed in the next section.


From this discussion and the examples of Figure \ref{fig:topology_condition}, we expect
that the possibility to define finite heat currents in a given circuit is related
to its topology. Indeed, this intuition is confirmed in the next section, where it is shown
that the quantities $\meann{\dot Q_r}$, as defined in Eq. \eqref{eq:def_local_heat},
are always well behaved if and only if, given a normal
tree of the graph of the circuit, there are no fundamental cut-sets associated to
it containing link resistors and twig resistors simultaneously (i.e, if
$Q_\text{RR}=0$ in Eq. \eqref{eq:block_Q}). In that case,
Eqs. (\ref{eq:local_heat}) and (\ref{eq:def_local_heat}) coincide.

\subsection{Fokker-Planck equation for the circuit state and heat currents}

Each resistor was modeled as an ideal resistor in series with a random
voltage noise. If $j_r$ is the instantaneous current flowing through the resistor,
then the rate of energy dissipation in it is
\begin{equation}
\dot Q_r = j_r (v_r + \Delta v_r),
\end{equation}
where $v_r = R_r j_r$ is the voltage drop in the ideal resistor and
$\Delta v_r$ is the random voltage. We want to express the quantity $\dot Q_r$ in
terms of the state $x$ of this circuit. For this, we first write it as
\begin{equation}
\dot Q_r =
\left[
\begin{array}{cc}
j_{R_l} & v_{R_t}
\end{array}
\right]
\Pi_r R
\left(
\left[
\begin{array}{c}
j_{R_l} \\ v_{R_t}
\end{array}
\right]
+
\left[
\begin{array}{c}
R_l^{-1} \Delta v_{R_l} \\ \Delta v_{R_t}
\end{array}
\right]
\right),
\end{equation}
where $\Pi_r$ is the projector associated with the $r$-th resistor appearing
in Eq. (\ref{eq:def_Cr}), and $R$ is the matrix of resistances defined in
Eq. (\ref{eq:def_matrix_R}). We can now use Eq. (\ref{eq:resistor_variables_stoch})
to eliminate the variables $j_{R_l}$ and $v_{R_t}$. For ease of notation, in the
following we will not consider the terms associated to the sources of the circuit,
as we know that they can only affect the mean values of the current and voltages and
we are only interested in the stochastic contributions to the heat currents.
In this way, after some manipulations, we obtain the following expression:
\begin{widetext}
\begin{equation}
\begin{split}
\dot Q_r = \: &x^T \mathcal{H} \mathcal{M}_{d}^T \alpha^T \Pi_r R \alpha \mathcal{M}_{d} \mathcal{H} x \: - \:x^T \mathcal{H} \mathcal{M}_{d}^T \alpha^T \Pi_r R \alpha
\left[
\begin{array}{c}
- \Delta v_{R_l} + Q_\text{RR} \Delta v_{R_t} \\
 Q_\text{RR} R_l^{-1} \Delta v_{R_l} + R_t^{-1} \Delta v_{R_t}
\end{array}
\right] \\
&+
\left[\!
\begin{array}{cc}
- \Delta v_{R_l}^T \!+\! \Delta v_{R_t}^T Q_\text{RR}^T, 0
\end{array}
\right]
 \alpha^t \Pi_r R \alpha
\left[
\begin{array}{c}
0 \\
Q_\text{RR} R_l^{-1} \Delta v_{R_l} \!+\! R_t^{-1} \Delta v_{R_t}
\end{array}
\!\!\right].
\label{eq:stoch_Qr_full}
\end{split}
\end{equation}
\end{widetext}
Notice that the last term in this expression is quadratic in the noise variables.
As a consequence, it will give a divergent contribution to $\meann{\dot Q_r}$
in the white noise limit, proportional to $\delta(0)$.
As we already mentioned, the physical origin of this divergence is the
possibility of direct heat transport at arbitrarily high frequencies between
resistors at different temperatures. We stress that this problem only arises
with regard to the definition of local heat currents, while the state of the
circuit $x$ and the total heat rate $\dot Q$ are always well behaved
quantities (see below). A solution to this problem would be to give a more realistic description
of the resistive thermal noise, associating to each resistor a spectral density $J_r(\omega)$
vanishing for large frequencies, such that its noise spectrum is
given by $S_r(\omega) = (R k_b T_r/\pi) J_r(\omega)$. However, this
is equivalent to appropriately `dressing' a white-noise resistor with inductors
and/or capacitors that can be considered part of the circuit\footnote{
This is fully analogous to well known `Markovian embedding'
techniques.} (a capacitor in parallel or a inductor in series
to a given resistor being the most simple options to filter out
high frequencies, see Figure \ref{fig:topology_condition}).
This observation hints at a relationship between the topology
of the circuit and the possibility of defining well behaved local heat currents.
In fact, we see that the quadratic terms in the noise
vanish if the matrix $\alpha^t \Pi_r R \alpha$ is block diagonal. In turn, from the
definition of $\alpha$ we can see that this happens if $Q_\text{RR} = 0$.
Thus, $Q_\text{RR} =0$ is a sufficient condition for the local heat currents
$\dot Q_r$ to be well behaved. In Appendix \ref{ap:topology_condition} it is
shown that this condition is also necessary. We also show in the same Appendix
that this condition can be restated in a way that does not make reference
to any tree.
To finish this discussion, we
note that since $\sum_r \Pi_r = \mathds{1}$ and $\alpha^t R \alpha$ is always block
diagonal (recall Eq. (\ref{eq:symm_alpha})), we see that the total
heat rate $\dot Q = \sum_r \dot Q_r $ is always well behaved in the white
noise limit, even if $Q_\text{RR} \neq 0$.

Thus, assuming $Q_\text{RR}=0$, the local heat currents are simplified to:
\begin{equation}
\dot Q_r = x^T \mathcal{H} \mathcal{M}_{d}^T \Pi_r R^{-1} \mathcal{M}_{d}
 \mathcal{H} x - x^T \mathcal{H} \mathcal{M}_{d}^T \Pi_r R^{-1} \eta,
\end{equation}
where $\eta$ is the vector of random voltages and currents defined in
Eq. (\ref{eq:def_eta}) and we used the fact that for $Q_\text{RR}=0$ we
have $\alpha = R^{-1}$. This equation is a Langevin equation for the heat
$Q_r$ that is of course coupled to the Langevin equation in Eq. (\ref{eq:langevin})
for the circuit state $x$. Their integration is to be performed according
to the Stratonovich procedure. As explained in \cite{gardiner2009} (Chapter 8),
in the white noise limit the corresponding stochastic dynamics can be described
by the following set of Ito differential equations:
\begin{equation}
dx = \mathcal{A}\:\mathcal{H}\:x\:dt
+\sum_r \sqrt{2k_b T_r} \: \mathcal{C}_r  \: dW,
\label{eq:ito_x}
\end{equation}
and
\begin{equation}
\begin{split}
dQ_r \!=\!
\Tr[ (\mathcal{H} xx^T \mathcal{H} \!-\! k_b T_r \mathcal{H}) \mathcal{C}_r\mathcal{C}_r^T]  dt
 &\!-\! \sqrt{2k_b T_r} x^T \mathcal{H}\mathcal{C}_r  dW,
\end{split}
\label{eq:ito_Qr}
\end{equation}
where $dW$ is a vector of independent Wiener processes differentials. Taking
the mean value of Eq. \eqref{eq:ito_Qr} we recover Eq. \eqref{eq:local_heat} for
$\meann{\dot Q_r}$.
The Fokker-Planck equation for the joint probability distribution $P(x,Q_r)$
corresponding to the previous Ito differential equations reads:
\begin{align}
\frac{dP}{dt} =& - \Tr[\mathcal{A}\mathcal{H}] P - x^T \mathcal{H}\mathcal{A}^T \nabla P \nonumber \\
&- \Tr[ (\mathcal{H} xx^T \mathcal{H} \!-\! k_b T_r \mathcal{H}) \mathcal{C}_r\mathcal{C}_r^T] \: \partial_{Q_r} P \nonumber \\
&-2k_b T_r \left( \Tr[\mathcal{H} \mathcal{C}_r \mathcal{C}_r^T] \partial_{Q_ r} P + x^T \mathcal{H} \mathcal{C}_r \mathcal{C}_r^T \: \nabla (\partial_{Q_ r} P)\right) \nonumber \\
&+k_b T_r \: x^T \mathcal{H} \mathcal{C}_r \mathcal{C}_r^T \mathcal{H} x \: \partial^2_{Q_r} P \nonumber \\
&+\sum_{r'} k_b T_{r'} \nabla^T \mathcal{C}_{r'}\mathcal{C}_{r'}^T \nabla P,
\label{eq:full_FP}
\end{align}
where $\nabla$ is the nabla operator with respect to the variables $x$. The
corresponding equation for the reduced probability distribution $p(x) = \int dQ_r \: P(x,Q_r)$
is just
\begin{equation}
\begin{split}
\frac{dp}{dt} =& - \Tr[\mathcal{A}\mathcal{H}] p
- x^T \mathcal{H}\mathcal{A}^T \nabla p
+\sum_{r} k_b T_{r} \nabla^T \mathcal{C}_{r}\mathcal{C}_{r}^T \nabla p.
\label{eq:redFP}
\end{split}
\end{equation}
Equation \eqref{eq:full_FP} allows to analyze the full statistics of the heat
currents. Indeed, different integrated and detailed fluctuation theorems can be
derived for this kind of linear and in general underdamped stochastic systems
\cite{murashita2016, jakvsic2017, damak2019}, valid for finite time protocols
or asymptotic steady states. However, in this article we will focus only on
the behaviour of the mean values.

\subsection{Entropy production}
We consider the continuous Shannon entropy
\begin{equation}
S = -k_b \int dx \: p(x) \log(p(x))
\end{equation}
associated to the distribution $p(x)$ and the total entropy production rate
\begin{equation}
\dot \Sigma = \frac{dS}{dt} + \sum_r \frac{\meann{\dot Q_r}}{T_r}.
\end{equation}
Using Eq. \eqref{eq:redFP} one can show that $\dot \Sigma$ is non-negative:
\begin{equation}
\dot \Sigma = \sum_r \frac{1}{T_r} \int dx \: p(x,t)\:
j_r(x,t)^T \mathcal{C}_r\mathcal{C}_r^T j_r(x,t) \geq 0,
\label{eq:second_law}
\end{equation}
where $j_r(x,t) = \mathcal{H}(t) x + k_b T_r \nabla \log(p(x,t))$. Eq. \eqref{eq:second_law}
is the Second Law of thermodynamics for the circuit.

This total entropy production can be decomposed as a sum of adiabatic and non-adiabatic contributions\cite{esposito2010,
van2010, esposito2010pre, spinney2012}, as shown in Appendix \ref{ap:ad_nonad_decomp}.
The adiabatic contribution is positive definite and for time independent circuits
it is the only non vanishing contribution for large times. In general the non-adiabatic
contribution can have any sign, but for overdamped circuits
(for example, circuits with no capacitors or no inductors)
it is also positive definite. If the circuit is time independent,
then this contribution equals $-k_b$ times the time derivative of the relative entropy $H(p|p_\text{st})$
between the instantaneous state $p(x,t)$ and the stationary one $p_\text{st}(t)$
(which for linear circuits is unique, although it might depend on the initial conditions).
Thus, for time independent circuits $H(p|p_\text{st})$ is a always decreasing Lyapunov function.
For underdamped circuits a third non-adiabatic term appears which, at variance with the
previous two, is not positive definite. It is related to the change in $H(p|p_\text{st})$
due to a conservative flow in phase space, and vanish identically in isothermal
conditions (when $p_\text{st}$ is an equilibrium state). These findings are
analogous to the results of \cite{spinney2012}.

\section{A simple circuit-based machine}
\label{sec:example}

\begin{figure}
\includegraphics[scale=.19]{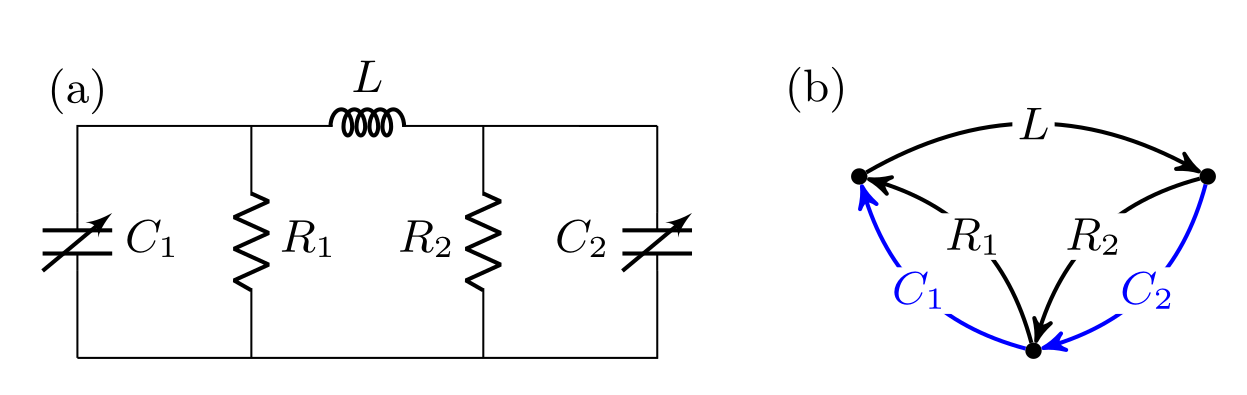}
\caption{(a) Two variable RC parallel circuits at possibly different
temperatures coupled by an inductor. (b) Graph of the circuit. The only normal
tree is shown in blue.}
\label{fig:example}
\end{figure}

We now illustrate how our formalism can be used to design thermodynamic
machines made of RLC circuits. External driving on the circuit
allows to implement thermodynamical cycles that might extract work from
a thermal gradient (non-autonomous heat engine), or  extract heat from some
resistors (non-autonomous refrigerator). We illustrate the basic techniques
by considering a minimal circuit which can work both as an engine or a refrigerator.
The circuit is shown in Figure \ref{fig:example} and
consists of two parallel RC circuits coupled by an inductor.
The capacitances in each RC circuit can be driven externally,
for example by changing in time the distances between the plates of each capacitor,
or, more practically, using varicap diodes. The circuit has no loops
consisting only of capacitors or cutsets of all inductors, and therefore
the previous formalism can be directly applied. We first analyze the simplest
case of regular heat conduction for constant parameters and different temperatures.
Then we show that the capacitances in the
circuit can be driven in time in order to cool (i.e, extract heat) from one
of the resistors, while dumping the extracted energy into the other one.
Similar circuits were analyzed before in \cite{ciliberto2013, ciliberto2013b, karimi2016}.

We begin by describing the circuit by the procedure of Section \ref{sec:description_circuits}. The state of the circuit is encoded in the
vector $x=(q_1, q_2, \phi)^T$, where $q_i$ is the charge in the capacitor $C_i$
and $\phi$ is the magnetic flux in $L$. According to the edge orientations of
Fig. \ref{fig:example}-(b), the cutset matrix associated to the normal tree is specified
by
\begin{equation}
Q_\text{link} =
\begin{array}{c|ccc}
& R_1 & R_2 & L\\
\hline
\textcolor{blue}{C_1} & 1 & 0 & -1\\
\textcolor{blue}{C_2}  & \undermat[15pt]{$Q_\text{CR}$}{ 0 & 1 &}  \!\undermat[15pt]{$Q_\text{CL}$}{-\!1}\\
\end{array},
\vspace{.5cm}
\end{equation}
and from this we can construct the matrix $\mathcal{M}_{c}$ and
$\mathcal{M}_{d}$:
\begin{equation}
\mathcal{M}_{c} = \left[
\begin{array}{cc|c}
0 & 0 & 1\\
0 & 0 & 1\\
\hline
-1 & -1 & 0\\
\end{array}
\right]
\qquad
\mathcal{M}_{d} = \left[
\begin{array}{cc|c}
-1 & 0 & 0\\
0 & -1 & 0\\
\end{array}
\right].
\end{equation}
Also, in this case $\alpha=R^{-1}=\text{diag}(1/R_1, 1/R_2)$ and therefore:
\begin{equation}
\mathcal{A}(t) = \mathcal{M}_{c} - \mathcal{M}_{d}^T
\bm{\alpha}\mathcal{M}_{d} =
\left[
\begin{array}{cc|c}
-R_1^{-1} & 0 & 1\\
0 & -R_2^{-1} & 1\\
\hline
-1 & -1 & 0\\
\end{array}
\right].
\end{equation}
Finally, we have $\mathcal{H} = \text{diag}(C_1, C_2, L)^{-1}$,
$\mathcal{C}_1\mathcal{C}_1^T = \text{diag}(R_1^{-1},0,0)$ and
$\mathcal{C}_2\mathcal{C}_2^T = \text{diag}(0,R_2^{-1},0)$.

\subsection{Heat conduction}
Given the above matrices, we can readily solve Eq. (\ref{eq:cov_matrix_st})
to find the stationary covariance matrix (for time independent parameters).
The solution is particularly simple in the symmetric case in which $R_1 = R_2 = R$
and $C_1 = C_2 = C$. It reads
\begin{equation}
\sigma = k_b \bar T \mathcal{H}^{-1} + \frac{k_b \Delta T}{2} \frac{CL}{CR^2+L}
\left[
\begin{array}{ccc}
1 & 0 & -R\\
0 & -1 & R\\
-R & R & 0\\
\end{array}
\right],
\end{equation}
where $\bar T = (T_1 + T_2)/2$ and $\Delta T = T_1 - T_2$. Similar results can be
found in \cite{ciliberto2013, ciliberto2013b}. Thus, the first term
in the previous expression is just the equilibrium covariance matrix
corresponding to the mean temperature $\bar T$. By examining the second term, we see
that a temperature bias will establish correlations between the capacitors
and the inductor, but not between the capacitors themselves.
Since this circuit is not forced by voltage or current sources, the mean values
of voltages and currents in any branch will vanish in the stationary state.
Then, introducing the
previous expression for the covariance matrix in Eq. (\ref{eq:local_heat}) for
the rate of heat dissipation in each resistor, we obtain:
\begin{equation}
\meann{\dot Q_1} = -\meann{\dot Q_2}
= -\frac{k_b \Delta T}{2} \frac{\tau_d}{\tau_d^2 + \tau_0^2},
\end{equation}
where we have introduced the two characteristic timescales of the circuit,
$\tau_d=RC$ and $\tau_0 = \sqrt{LC}$, respectively associated to the dissipation rate and the free oscillations period.
The entropy production can be computed from the above heat rates and reads:
\begin{equation}
\begin{split}
\meann{\dot \Sigma} &= \sum_r \frac{\meann{\dot Q_r}}{T_r} =
\meann{\dot Q_1} \left(\frac{1}{T_1}-\frac{1}{T_2}\right)\\
&=\frac{k_b (\Delta T)^2}{2 T_1 T_2}
\frac{\tau_d}{\tau_d^2 + \tau_0^2}
\geq 0.
\end{split}
\end{equation}
Thus, in this case, the only dissipation in this circuit corresponds
to static heat conduction from the hot to the cold reservoir.

\subsection{Cooling cycle}

\begin{figure*}
  \includegraphics[width=\textwidth]{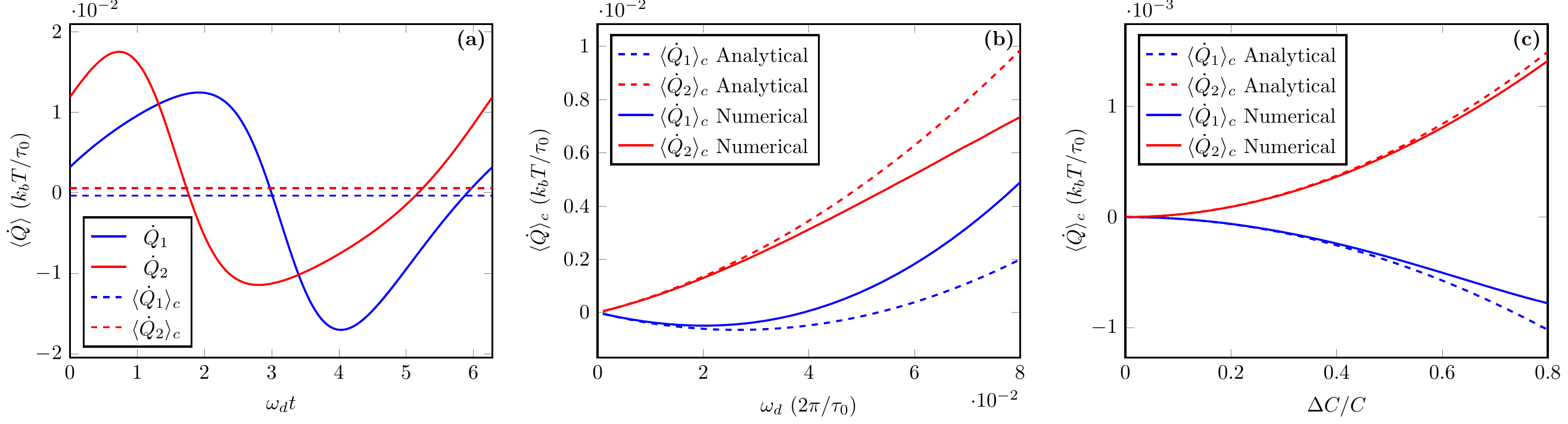}
  \caption{
  (a) Asymptotic cycle of the heat currents for $\Delta C/C = 1/2$ and
  $\omega_d/(2\pi) = 10^{-2}/\tau_d$ (dashed lines indicate cycle averages).
  (b) Average heat currents versus driving frequency for $\Delta C/C = 0.5$.
  (c) Average heat currents versus driving strength for $\omega_d/(2\pi) = 10^{-2}/\tau_d$.
  For all cases we took $\theta = \pi/2$ and $T_1 = T_2 =T$.}
  \label{fig:example_results}
\end{figure*}

We now turn to analyze the more interesting situation in which the two
resistors are at the same temperature $T_1 = T_2 = T$ and the capacitors
are driven periodically in time such that $C_1(t) = C + \Delta C \cos(\omega_d t)$ and
$C_2(t) = C + \Delta C \cos(\omega_d t + \theta)$. Thus, both capacitances
are driven with the same angular frequency $\omega_d$ and the same amplitude $\Delta C$,
but with some fixed phase difference $\theta \in (-\pi,\pi)$. The matrices describing the circuit are
the same as before except for the energy matrix $\mathcal{H}(t)$ that now depends
on time. Since it is periodic, it can be decomposed as a Fourier series:
\begin{equation}
\mathcal{H}(t) = \sum_{k=-\infty}^{+\infty} \mathcal{H}_k \: e^{ik\omega_d t}.
\end{equation}
To lower order in $\Delta C$ we have only three Fourier components:
$\mathcal{H}_0 = \text{diag}(C,C,L)^{-1}$ and
$\mathcal{H}_{\pm1} = -\Delta C/(2C^2) \: \text{diag}(1,e^{\pm i \theta},0)$.
We focus in regimes when a stable stationary state is reached. We note
that this is not always the case due to the phenomenon of parametric resonance \cite{poulin2008}.
However, if there is a stable stationary state it will be such that the mean
values of voltages and currents vanish, while the covariance matrix is periodic
with the same period as the driving. Thus, we can decompose it as
\begin{equation}
\sigma(t) = \sum_{k,k'=-\infty}^{+\infty} \sigma_{k,k'} \: e^{i(k-k')\omega_d t}.
\label{eq:cov_matrix_fourier}
\end{equation}
Inserting the previous two Fourier decompositions into Eq. (\ref{eq:cov_matrix_evol}),
we obtain an algebraic equation from which it is possible to obtain the
coefficients $\sigma_{k,k'}^2$ in terms of $\mathcal{H}_0$ and $\mathcal{H}_{\pm1}$.
This technique is explained in Appendix \ref{ap:gen_lyapunov}.
Once the Fourier components $\sigma_{k,k'}^2$ have been
determined, we compute the \emph{average heat and work rates per cycle}.
Explicitly, we consider the quantity
\begin{equation}
\meann{\dot{X}}_c \equiv \lim_{t\to \infty}\frac{\omega_d}{2\pi} \int_t^{t+\frac{2\pi}{\omega_d}}
\meann{\dot X} \: dt,
\end{equation}
where $X$ stands for $Q_1$, $Q_2$, $W$ or $E$. We note that $\meann{\dot{E}}_c = 0$
since the asymptotic state of the system is periodic. Thus, averaging the
balance of energy (Eq. \eqref{eq:energy_balance}) during a cycle we obtain
\begin{equation}
\meann{\dot{W}}_c = \meann{\dot{Q}_1}_c + \meann{\dot{Q}_2}_c,
\end{equation}
where $\meann{\dot{W}}_c$ is the average rate of work corresponding to the external
driving, which is the only source of work in this case.

As shown in Appendix \ref{ap:gen_lyapunov}, to lower order in $\Delta C$ and to
second order in $\omega_d$, $\meann{\dot{W}}_c$ is given by
\begin{equation}
\meann{\dot{W}}_c =
k_b T
\omega_d^2 \tau_d \left(\! \frac{\Delta C}{2C} \! \right)^2
\frac{  \tau_d^2 (1+\cos(\theta))/2 + \tau_0^2}{\tau_d^2 + \tau_0^2}
+ \mathcal{O}(\omega_d^ 3),
\end{equation}
while the average heat currents are
\begin{equation}
\meann{\dot{Q}_{1/2}}_c =
\mp \frac{k_b T \: \omega_d}{2}  \left(\! \frac{\Delta C}{2C} \! \right)^2
\! \frac{\tau_d^4 \sin(\theta)}{(\tau_d^2 + \tau_0^2)^2}
+ \frac{\meann{\dot{W}}_c}{2}  + \mathcal{O}(\omega_d^ 3).
\label{eq:av_heat_ex}
\end{equation}
We then see that the rate of heat pumping from one resistor to the other (the first
term in Eq. \ref{eq:av_heat_ex}) is proportional to $\omega_d$, while the
rate at which work is performed by the driving (or equivalently, the total dissipated heat rate)
scales as $\omega_d^2$. Also, the pumping of
heat is maximized and the dissipated work minimized for $\theta = \pm \pi/2$.
This is natural since in this case the left/right asymmetry induced by the driving
is maximum. The cooling efficiency or `Coefficient of Performance' is
\begin{equation}
\text{CoP} = \frac{|\meann{\dot{Q}_1}_c|}{\meann{\dot W}_c}
= \frac{\tau_d/\omega_d}{\tau_d^2 + \tau_0^2} \: \frac{\sin(\theta)}{1+\cos(\theta)+2(\tau_0/\tau_d)^2}
-\frac{1}{2}.
\end{equation}
We note that in this isothermal case the $\text{CoP}$ is not bounded by the Second law,
and in fact diverges in the quasistatic limit $\omega_d \to 0$.
There is a maximum driving frequency such that cooling is not possible above
it (in the considered regime of low $\omega_d$ and $\Delta C$).
It corresponds to $\text{CoP} = 0$ and for $\theta \in [0,\pi]$ reads
\begin{equation}
\omega_d^\text{max} = \frac{2\tau_d}{\tau_d^2+\tau_0^2} \:
\frac{\sin(\theta)}{1+\cos(\theta)+2(\tau_0/\tau_d)^2}.
\label{eq:wd_max}
\end{equation}
There is also an optimal frequency $\omega_d^\text{opt}$, in the sense that the heat extracted from
one of the resistors is maximized. We can obtain it by optimizing $\meann{\dot Q_1}_c$
in Eq. \eqref{eq:av_heat_ex} with respect to $\omega_d$, and in this way we find
that $\omega_d^\text{opt} = \omega_d^\text{max}/2$.

An intuitive understanding of how the cooling effect is achieved
can be obtained by analyzing Eq. (\ref{eq:local_heat}) for the
heat currents. We note that the stochastic contribution
is proportional to the difference between the energy
contained in the circuit elements connected to a given resistor,
and the energy they would have if they were
 in thermal equilibrium at the
temperature of that resistor.
For example, for the circuit of Figure
\ref{fig:example}, we have
$\meann{\dot Q_1} = [\mean{q_1^2}\! (t)/(2C_1) - k_b T_1/2]/(R_1C_1/2)$.
Thus, we see that $\meann{\dot Q_1}$ will be positive whenever
the average energy contained in $C_1$, $\mean{q_1^2}\!(t)/(2C_1)$, is
larger that the corresponding to equilibrium, $k_b T_1/2$. Therefore,
to achieve cooling of $R_1$ we require the variance
$\mean{q_1^2}\!(t)$ to be, on average during a cycle, lower than its
equilibrium value $k_b T_1 C_1$. This reduction in the variance in
one degree of freedom with respect to its equilibrium value is the
classical analogue to the well known concept of quantum squeezing
\cite{rugar1991, natarajan1995}.
In quantum electronic and quantum optical setups,
squeezing is a useful resource for metrology and it is usually achieved
by means of some form of parametric driving \cite{yariv1967,anisimov2010, pirkkalainen2015, wollman2015}.
Here we see its additional role as a thermodynamic resource, which supports the
already mentioned observation that parametric amplifiers can also
be employed as refrigerators \cite{bergeal2010}. Thus, an optimal cooling
strategy is one in which a highly squeezed state is created and maintained
in a dissipative and non-equilibrium environment. Finally, we mention that
the methods of Appendix \ref{ap:gen_lyapunov} can be employed to numerically optimize
thermal cycles.


\subsection{Non-isothermal case}

If the resistor temperatures are different the heat currents are
\begin{equation}
\begin{split}
&\meann{\dot Q_{1/2}}_c =
\mp \frac{k_b \Delta T}{2} \frac{\tau_d}{\tau_d^2 + \tau_0^2}\\
&\mp \frac{k_b \Delta T}{2} \frac{\tau_d^3}{\tau_d^2 + \tau_0^2}
\left(\!\frac{\Delta C}{2C}\!\right)^2
\frac{\cos(\theta)(2\tau_d^2+\tau_0^2)-2\tau_d^2-3\tau_0^2}{(\tau_d^2 + \tau_0^2)^2}\\
& \mp \frac{k_b T_{1/2} }{2} \: \omega_d
\left(\!\frac{\Delta C}{2C}\!\right)^2 \frac{\tau_d^4 \sin(\theta)}{(\tau_d^2 + \tau_0^2)^2}
+ \mathcal{O}(\omega_d^2).
\end{split}
\end{equation}
In contrast to the isothermal case, the term of second order in $\omega_d^2$
is too involved to be shown here. The first term corresponds to regular heat
conduction in response to the thermal gradient. The second term is
a correction to the regular heat conduction due to the driving, while the
third term describes the pumping of heat.  We consider the case in which $T_1 < T_2$
(then, $\Delta T=T_1-T_2 <0$) and analyze the conditions under which it is
possible to extract heat from $R_1$. The pumping of heat out of $R_1$ is, as before,
optimized for $\theta = \pi/2$. From the previous equation we see that in general
the driving frequency must be above a minimum value in order for the heat pumping
to overcome the heat conduction imposed by the thermal gradient. Thus, we will
have effective cooling of $R_1$ only if $\omega_d > \omega_d^\text{min}$.
For $\theta = \pi/2$  this minimum cooling frequency $\omega_d^\text{min}$ reads
\begin{equation}
\omega_d^\text{min} = \frac{|\Delta T|}{T_1}
\left[
\left(\frac{2C}{\Delta C}\right) \frac{\tau_d^2 + \tau_0^2}{\tau_d^3}
-
\frac{2\tau_d^2+3\tau_0^2}{\tau_d(\tau_d^2+\tau_0^2)}
\right],
\label{eq:wd_min}
\end{equation}
and we can write the heat rate $\meann{\dot{Q}_1}$ as:
\begin{equation}
\meann{\dot{Q}_1}_c = \frac{k_b T_1}{2}  (\omega_d^\text{min} - \omega_d)
\left(\!\frac{\Delta C}{2C}\!\right)^2 \frac{\tau_d^4}{(\tau_d^2 + \tau_0^2)^2}
+ \mathcal{O}(\omega_d^2).
\label{eq:example_heat_noniso}
\end{equation}
The previous considerations do not take into account the terms of second order
in $\omega_d$. From the expression of the heat currents in the isothermal case,
Eq. (\ref{eq:av_heat_ex}), we know that these corrections correspond to heating
and establish a maximum driving frequency $\omega_d^\text{max}$ such that
cooling is not possible above it, Eq. (\ref{eq:wd_max}). Thus, for
cooling to be possible at all we need that $\omega_d^\text{min} < \omega_d^\text{max}$,
which imposes a condition on the temperature difference.

We now turn to analyze the total heat rate, or work rate. Up to second order
in $\omega_d$ it is given by the following expression:
\begin{equation}
\begin{split}
&\meann{\dot W}_c =
-\frac{k_b \Delta T}{2} \omega_d \left(\!\frac{\Delta C}{2C}\!\right)^2
\! \frac{\tau_d^4 \sin(\theta)}{(\tau_d^2 + \tau_0^2)^2}\\
& + k_b \bar{T} \omega_d^2 \tau_d \left(\!\frac{\Delta C}{2C}\!\right)^2
\! \frac{\tau_d^2(1+\cos(\theta))/2+\tau_0^2}{\tau_d^2+\tau_0^2}
+ \mathcal{O}(\omega_d^3).
\end{split}
\label{eq:example_work_noniso}
\end{equation}
Note that that if $\Delta T \neq 0$ then to lower order in $\omega_d$ the
average work rate can be positive or negative, depending on the value of $\theta$.
This two cases correspond to the device working as a refrigerator or a (non-autonomous) heat engine,
respectively. From Eqs. \eqref{eq:example_heat_noniso} and \eqref{eq:example_work_noniso}
it follows that the cooling efficiency in this case, to lower order in $\omega_d$, is:
\begin{equation}
\text{CoP} = \left(1- \frac{\omega_d^\text{min}}{\omega_d} \right) \frac{T_1}{T_2 - T_1},
\end{equation}
which is of course bounded by the Carnot efficiency $\text{CoP}_\text{Carnot} = T_1/(T_2-T_1)$.

\subsection{Exact numerical results}

The previous analytical results for the cooling protocol are limited to low
driving amplitude and frequency. In order to assess their validity in that regime
and to study the behaviour of the system away from it, we numerically compute
the heat currents. For this we integrate the differential equation for the time
evolution of the covariance matrix (Eq. (\ref{eq:cov_matrix_evol})). Then we
compute the instantaneous expected values for the heat currents via Eq.
(\ref{eq:local_heat}), and obtain their averages during a cycle for sufficiently
long times.
For the numerical evaluation we consider $\tau_d = \tau_0$ and take this
quantity as the unit of time. As an example we
show in Figure \ref{fig:example_results}-(a) the long time oscillations of the heat currents,
as well as their averages, for an isothermal setting and the following
driving parameters: $\Delta C/C = 1/2$, $\omega_d = 10^{-2}\: 2\pi/\tau_d$
and the optimal phase difference of $\theta = \pi/2$.
The analytical and numerical results are compared in Figure \ref{fig:example_results}-(b)
for a fixed driving amplitude ($\Delta C/C = 0.5$) and increasing driving frequency,
while the temperatures are the same and the phase difference is the optimal.
We see that the analytical expressions indeed match the numerical results in the
low driving frequency regime. We also see that there is, as expected from the
theoretical analysis, a maximum
driving frequency $\omega_d^\text{max}$ such that both heat currents are positive
if $\omega_d > \omega_d^\text{max}$. However, the analytical results overestimate
the value of $\omega_d^\text{max}$.
Analogously, we show in Figure \ref{fig:example_results}-(c) the heat currents for
fixed driving frequency ($\omega_d/(2\pi) = 10^{-2}/\tau_d$) and increasing
driving amplitude. Again, we see that for low driving amplitude the analytical
expressions correctly describe the numerical results.

\begin{figure}
  \includegraphics[scale=1]{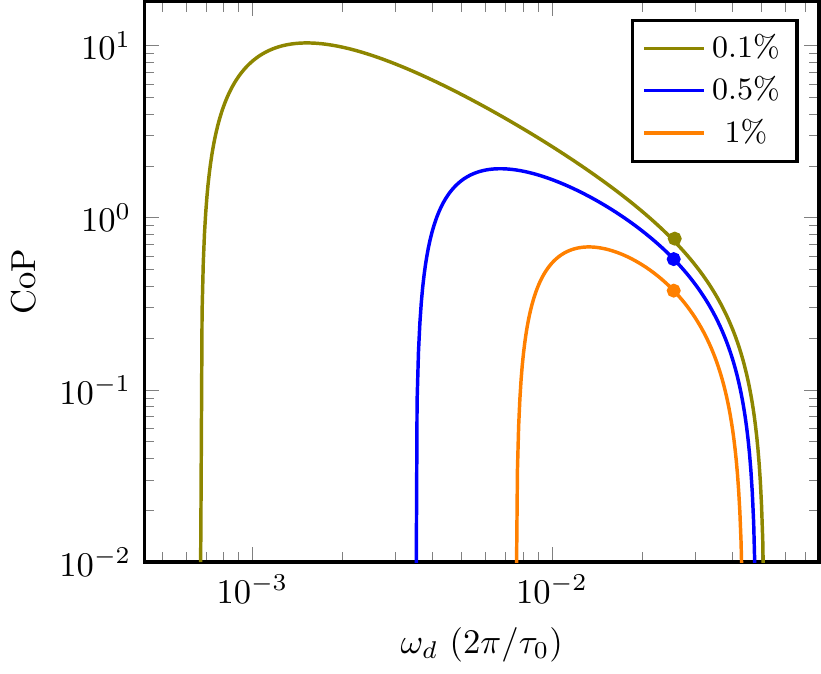}
  \caption{
  Coefficient of Performance for different values of
  opposing thermal gradient $|\Delta T|$
  ($0.1\%$, $0.5\%$ and $1\%$ of $T_1$). The points in each curve indicate the driving frequency for which the cooling power is maximum. The parameters in this case are $\Delta C/C = 0.6$ and $\tau_d = 3 \tau_0$.
  }
  \label{fig:CoP}
\end{figure}
Finally, in Figure \ref{fig:CoP} we show the Coefficient of Performance as a function of the driving frequency for different
values of $\Delta T$. We see that cooling is possible only in a clearly defined range of driving frequencies.
The lower limit of this range increases with $|\Delta T|$, in
accordance with Eq. \eqref{eq:wd_min}, while the upper limit displays
a weaker dependence on the same parameter. The points in each curve correspond to maximum cooling power, which is attained in all the cases at a frequency close to the optimal driving frequency of the isothermal case, $\omega_d^\text{opt}$. In all cases the maximum CoP is only about $1\%$ of the ideal value, $\text{CoP}_\text{Carnot}$. Thus, we see that this simple cooling scheme
can only withstand small thermal gradients and operate at low efficiencies. More complex
schemes are expected to improve these figures of merit, although they will probably
share some of the properties of this elementary example, like the existence
of minimum and maximum driving frequencies for cooling.

\section{Quantum Johnson-Nyquist noise}
\label{sec:quantum}

We now turn to the low temperature regime of our theory.
If the typical temperature in the circuit is low enough that the
thermal energy $k_b T$ starts to be comparable to the quantum
of energy $\hbar \omega$ at the relevant frequencies $\omega$, then the
quantum nature of the noise in each resistor must be taken into account.
One approach to work in this regime is to construct a quantum model
of the circuit, assigning quantum mechanical operators to the charge and
flux degrees of freedom in it (or a proper combination of them),
satisfying the usual commutation relations. The dissipation and diffusion effects
induced by the resistors are typically introduced using the Caldeira-Legget model
for quantum Brownian motion \cite{caldeira1983}. This is certainly
the way to go if one is interested in having access to the full quantum state of
the circuit (for modern treatments,
see \cite{devoret1995,solgun2015,clerk2010,parra2019,burkard2004,girvin2011}).
However, this cannot be directly done for any circuit, since the
canonical quantization procedure requires the detailed specification of stray or parasitic
capacitances and inductances \cite{devoret1995}. As an example we can take
the overdamped circuit of Fig. \ref{fig:topology_condition}-(c). This circuit
has well behaved classical dynamics and heat currents, but cannot be directly quantized
since it is missing inertial degrees of freedom. In other words, it is not possible
to define canonical `momentum' coordinates that are conjugate of the capacitor charges (there
are no kinetic energy terms \cite{parra2019}). To do that, one must give a more
detailed description specifying stray inductances.
Of course, any actual component in a real circuit is characterized by an impedance
that is never purely resistive, capacitive or inductive.
At a fundamental level we could consider elementary components which are always a combination
of a single inductor and a single capacitor (resistors are then modeled as infinite arrays
of them). Any circuit constructed in this way can be directly quantized,
and based on this quantization the low temperature behaviour can be studied.
Circuits like the one in Fig. \ref{fig:topology_condition}-(c) result from having
disregarded the inductive component in the impedance of the actual capacitors,
which anyway might be an excellent approximation for practical purposes.
However, this procedure of ignoring some degrees of freedom in the circuit is
problematic, as we can see by considering the simple example of a series RLC circuit. The relevant
frequency scales in this case are given by the dissipation rate $\gamma = R/L$,
the oscillation frequency $\omega_0 = 1/\sqrt{LC}$, and the thermal frequency
$\omega_\text{th} = k_bT/\hbar$. 
Under the high temperature condition $\omega_\text{th} \gg \omega_0, \gamma$,
the equilibrium charge variance is $\mean{q^2} \simeq k_b T C$ and does not
depend on $L$.
This justifies the use of an overdamped RC model, where the inductance $L$ is neglected compared to $R^2C$,
in the high temperature regime.
In contrast, if we take the overdamped limit $L/(R^2C) \to 0$ for low temperatures we obtain that the equilibrium charge
variance diverges as $\mean{q^2} \propto L^{-1}$ while the flux variance vanishes
as $\mean{\phi^2} \propto L$ (these scalings are obtained by keeping $\omega_0$ constant,
otherwise we have $\mean{q^2} \propto L^{-1/2}$ and $\mean{\phi^2} \propto L^{1/2}$).
Thus, in general the covariance matrix of the circuit state is not well defined in
this limit, and to compute it one needs to have precise information about the value
of $L$. In spite of this, we will see in the following that the heat currents are
actually well defined in this kind of overdamped limits, and can be directly evaluated
from the overdamped description of circuits even at low temperatures.

To show this we put forward a semiclassical approach that is based on the
same stochastic equations of motion of Eq. \eqref{eq:langevin}, where now
the noise variables $\xi(t)$ are not white anymore and display a quantum spectrum.
Then, if $\xi_r$ is the adimensional noise variable associated with the $r$-th resistor,
we must consider the following power spectrum:
\begin{equation}
\begin{split}
\mathcal{S}_{r,r'}(\omega) &= \frac{1}{2\pi} \int_{-\infty}^{+\infty}\!d\tau \: e^{-i\omega \tau} \meann{\xi_r(t)\xi_{r'}(t+\tau)} \\
&= \frac{\delta_{r, r'}}{2\pi} \frac{\hbar\omega}{k_b T_r}\:(N_r(\omega) + 1/2),
\end{split}
\label{eq:quantum_spectrum}
\end{equation}
where
$N_r(\omega) = (e^{\hbar\omega/(k_b T_r)}-1)^{-1}$
is the Planck's distribution at temperature $T_r$. Inverting the previous equation,
we can compute the correlation functions:
\begin{equation}
\meann{\xi_r(t)\xi_{r'}(t+\tau)} = \frac{\delta_{r,r'}}{2\pi} \int_{-\Lambda}^{+\Lambda} \!\!\!\!\! d\omega
\: e^{i\omega \tau}\:\frac{\hbar\omega}{k_bT_r} \:(N_r(\omega)+1/2), \\
\label{eq:quantum_corr}
\end{equation}
where $\Lambda$ is a high-frequency cutoff, that must be large compared to any other frequency
scale of the problem. 
The variables $x$ specifying
the circuit state remain classical (they are not promoted to quantum mechanical operators).
However, it is possible to show that for linear circuits that can
be directly quantized, the results obtained in this way are fully equivalent to those obtained by
a full quantization (under the Markovian approximation) \cite{schmid1982, freitas2017}.

\subsection{Covariance matrix and heat currents for quantum noise}
Since $\mean{\xi}(t)=0$, for linear circuits the equation of motion for the mean
values is still the fully deterministic one given in Eq. \eqref{eq:mean_values_evol}.
However, in the quantum low temperature regime, the
 differential equation for the covariance matrix, Eq. \eqref{eq:cov_matrix_evol},
 must be modified. The reason is that
the circuit state $x(t)$ at time $t$ will be in general correlated to $\xi(t)$ (i.e,
it will not be a non-anticipating function, and therefore the usual assumptions
of stochastic calculus underlying the derivation of Eq. \eqref{eq:cov_matrix_evol}
are not valid \cite{gardiner2009}). For stable systems it is still possible to obtain a simple
expression for the covariance matrix at large times.
To see this explicitly it is convenient to employ techniques based on the Green's
function of the circuit, like it is done in fully quantum mechanical models.
We start with the equation of motion for $y = x-\mean{x}$:
\begin{equation}
\frac{dy}{dt} = \mathcal{A}(t)\mathcal{H}(t) \: y +
\sum_r \sqrt{2k_b T_r} \:\mathcal{C}_r(t) \:  \xi(t).
\end{equation}
Given the initial value $y(0)$, the solution to this equation can be written as
\begin{equation}
y(t) = G(t,0) \: y(0) + \int_0^t \! d\tau \: G(t,\tau) \sum_r \sqrt{2k_b T_r} \: \mathcal{C}_r(\tau)  \: \xi(\tau),
\end{equation}
where the retarded Green's function $G(t,t')$ is defined as the solution of
\begin{equation}
\frac{d}{dt} G(t,t') - \mathcal{A}(t)\mathcal{H}(t) G(t,t') = \mathds{1} \delta(t,t'),
\label{eq:def_green}
\end{equation}
with $G(t,t') = 0$ for $t<t'$ (from this it follows that $G(t',t')=\mathds{1}$).
Then, the covariance matrix can be expressed as
\begin{widetext}
\begin{equation}
\begin{split}
 \sigma(t) = \meann{y(t)y(t)^T} &= G(t,0) \sigma(0) G(t,0)^T\\
&+ \int_0^t \! d\tau \sum_r \sqrt{2k_b T_r}\: \left[G(t,0) \mean{y(0)\xi^T(\tau)} \mathcal{C}_r(\tau)^T G(t,\tau)^T  + G(t,\tau) \mathcal{C}_r(\tau)\mean{\xi(\tau)y(0)^T} G(t,0)^T\right]\\
&+ \int_0^t \! d\tau \int_0^t \! d\tau' \sum_{r, r'} 2k_b \sqrt{T_r T_{r'}}\:
 G(t,\tau) \mathcal{C}_r(\tau)\mean{\xi(\tau)\xi(\tau')^T}  \mathcal{C}_{r'}(\tau')^T G(t,\tau')^T.
\end{split}
\label{eq:full_quantum_cov}
\end{equation}
\end{widetext}
The first term in this expression is just the deterministic evolution of the
fluctuations present in the initial state. The second term takes into account
the effect of the correlations between the circuit initial state and the
environmental noise. The last term, which for stable systems dominates the
long time behaviour, represents the diffusion induced by the environment.
We will assume in the following that the initial state is not correlated
in any way with the environmental noise, $\mean{y(0)\xi(\tau)^T} =0$, and
therefore the second term in the previous equation vanish. We will also
assume, for simplicity, that the resistances, and thus the matrices
$\mathcal{A}$ and $\mathcal{C}_r$, are constant. Then, taking the
time derivative of Eq. (\ref{eq:full_quantum_cov}) and using Eq. (\ref{eq:def_green}),
we obtain the following differential equation:
\begin{equation}
\begin{split}
\frac{d}{dt} \sigma(t) &= \mathcal{A}\mathcal{H}(t)\sigma(t) +
\sigma(t) \mathcal{H}(t)\mathcal{A}^T \\
&+\sum_r 2k_b T_r \: \left(\mathcal{I}_r(t) \: \mathcal{C}_r\mathcal{C}_r^T +
\mathcal{C}_r\mathcal{C}_r^T \: \mathcal{I}_r(t)^T\right),
\label{eq:cov_matrix_evol_quantum}
\end{split}
\end{equation}
where $\mathcal{I}_r(t)$ is the convolution between the Green's function $G(t,t')$
and the correlation function of resistor $r$:
\begin{equation}
\mathcal{I}_r(t) = \int_0^t d\tau \: G(t,t-\tau) \: \meann{\xi_r(0)\xi_r(\tau)}.
\label{eq:def_I}
\end{equation}
Eq. \eqref{eq:cov_matrix_evol_quantum} is the generalization for quantum
noise of Eq. \eqref{eq:cov_matrix_evol}, which is recovered
in the limit of high temperatures. To see this, we note that for high temperatures
$\meann{\xi_r(0)\xi_r(\tau)} \to \delta(\tau)$,
and therefore $\mathcal{I}_r(t) \to G(t,t)/2 = \mathds{1}/2$.

Based on these results, we can now derive an expression for the local
heat currents that, in contrast to Eq. (\ref{eq:local_heat}), is exact
and valid for arbitrary temperatures. In the quantum case the total heat rate
is also given by Eq. (\ref{eq:stoch_Q}):
$\dot{\mean{Q}} = \dot Q (\mean{x}, t) - \frac{1}{2}\Tr\left[\mathcal{H}\frac{d\sigma}{dt}\right]$.
However, this time we should replace $d\sigma/dt$ by Eq. (\ref{eq:cov_matrix_evol_quantum}).
Using this and the FD relation, we can write:
\begin{equation}
\dot{\mean{Q}} = \sum_r \mean{j_r}\! \mean{v_r}
+ \Tr\left[ (\mathcal{H}\sigma(t)\mathcal{H}
- 2 k_b T_r \mathcal{H}\mathcal{I}_r(t)
)\mathcal{C}_r \mathcal{C}_r^T\right].
\end{equation}
In analogy with the classical case, under the condition $Q_\text{RR}=0$,
we can identify the local heat currents as:
\begin{equation}
\meann{\dot{Q}_r} = \mean{j_r}\! \mean{v_r}+
 \Tr\left[ (\mathcal{H}\sigma(t)\mathcal{H}
- 2 k_b T_r \mathcal{H}\mathcal{I}_r(t)
)\mathcal{C}_r \mathcal{C}_r^T\right].
\label{eq:local_heat_quantum}
\end{equation}
This expression can be evaluated using the above equations for $\sigma$ and
$\mathcal{I}_r$. However, if we are only interested in the asymptotic heat currents,
under the assumption that the dynamics of the system is stable, we can express
them as frequency integrals that might be easier to compute,
and that also have a clear physical interpretation in terms of elementary transport processes.

\subsection{Asymptotic covariance matrix and heat currents}
 If the system is asymptotically stable, i.e, if
$G(t, t') \to 0$ for $|t-t'| \to \infty$, then the first two terms
in Eq. \eqref{eq:full_quantum_cov} can be neglected for sufficiently long times.
By expressing the correlations $\mean{\xi(\tau)\xi(\tau')^T}$ in terms of the power spectrum
$\mathcal{S}_{r,r'}(\omega)$ via inversion of Eq. (\ref{eq:quantum_spectrum}),
we can rewrite the last term in Eq. (\ref{eq:full_quantum_cov}) as:
\begin{equation}
\begin{split}
\sigma(t) = \frac{1}{\pi}\! \sum_r \! \int_{-\Lambda}^{+\Lambda}\!\!\!\!\!\!\!\! d\omega \: \hbar\omega
\:  \hat G(t,\omega) \mathcal{C}_r \mathcal{C}_r^T \hat G(t,\omega)^\dagger (N_r(\omega)\! +\!1/2),
\label{eq:sigma_fourier}
\end{split}
\end{equation}
where we have defined the following partial transform of the Green's function:
\begin{equation}
\hat G(t,\omega) = \int_0^t  d\tau \: e^{-i\omega (t-\tau)} \: G(t,\tau).
\label{eq:def_G}
\end{equation}
For circuits that can be directly quantized, Eq. \eqref{eq:sigma_fourier} is
equivalent to what one obtains from a fully quantum model of the network
and its environment under the Markovian approximation \cite{freitas2017}.

If the circuit parameters are periodically driven, $\hat G(t,\omega)$
has the useful property of being asymptotically periodic in time with the same
period as the driving, as shown in Appendix \ref{ap:periodic_driving}.
It also trivially satisfies $\hat G(t,\omega)^* = \hat G(t,-\omega)$,
a property that is sometimes used implicitly in the derivations below.
The convolution integral $\mathcal{I}_r(t)$ can also be expressed in
terms of $\hat G(t,\omega)$:
\begin{equation}
\mathcal{I}_r(t) = \frac{1}{2\pi k_b T_r} \int_{-\Lambda}^{+\Lambda}
d\omega \:\hbar\omega \:\hat G(t,\omega) (N_r(\omega)+1/2),
\label{eq:def_I_freq}
\end{equation}
Finally, we note that $\hat G(t,\omega)$ can be directly obtained by
solving its own evolution equation, that can be derived from Eq. (\ref{eq:def_green})
and reads
\begin{equation}
\frac{d}{dt} \hat G(t,\omega) = \mathds{1} -[i\omega - \mathcal{A} \mathcal{H}(t)] \hat G(t,\omega)
\label{eq:evol_G}
\end{equation}
with the initial condition $\hat G(t=0,\omega) = 0$.

Introducing Eqs. (\ref{eq:sigma_fourier}) and (\ref{eq:def_I_freq}) for $\sigma$
and $\mathcal{I}_r$ into Eq. (\ref{eq:local_heat_quantum}), we can write
the local heat currents as:
\begin{widetext}
\begin{equation}
\meann{\dot{Q}_r} = \mean{j_r}\! \mean{v_r} +\\
\frac{1}{\pi} \sum_{r'} \int_{-\Lambda}^{+\Lambda} d\omega \: \hbar\omega \:
\Tr\left[\left( \mathcal{H} \hat G(t,\omega) \mathcal{D}_{r'}
\hat G(t,\omega)^\dagger \mathcal{H} -\delta_{r,r'} \mathcal{H} \hat G(t,\omega) \right)
\mathcal{D}_{r} \right]
  (N_{r'}(\omega)+1/2),
\label{eq:local_heat_freq_1}
\end{equation}
\end{widetext}
where we introduced the shorthand definition $\mathcal{D}_r = \mathcal{C}_r \mathcal{C}_r^T$,
that we will employ in the following to simplify the notation.
The first term inside the trace, that is quadratic in $\hat G(t,\omega)$, is actually closely
related to the second one, which is linear in $\hat G(t,\omega)$.
We can see this by employing Eq. (\ref{eq:evol_G}) to compute the derivative
of $\hat G^\dagger \mathcal{H} \hat G$:
\begin{equation}
\frac{d}{dt}\!\!\left(\hat G^\dagger\mathcal{H}\hat G \right)
-\hat G^\dagger\frac{d\mathcal{H}}{dt}\hat G
-2\hat G^\dagger \mathcal{H} (\mathcal{A})_s \mathcal{H} \hat G
=
\mathcal{H}\hat G + \hat G^\dagger\mathcal{H}.
\label{eq:evol_GHG}
\end{equation}
Using this relationship and Eq. \eqref{eq:fd_relation}, it is possible to rewrite Eq. \eqref{eq:local_heat_freq_1}
in the following compact way:
\begin{equation}
\meann{\dot Q_r} =  \mean{j_r}\! \mean{v_r} + \sum_{r'} \int_{-\Lambda}^{+\Lambda} d\omega \: \hbar \omega \:
f_{r,r'}(t,\omega) \: (N_{r'}(\omega)+1/2),
\label{eq:local_heat_freq_2}
\end{equation}
where $f_{r,r'}(t, \omega)$ is a transfer function, specifying how the temperature
of resistor $r'$ affects the heat current of resistor $r$. For $r\neq r'$  it is
always positive and is given by:
\begin{equation}
f_{r,r'}(t,\omega) = \frac{1}{\pi} \Tr\left[\mathcal{H}(t)\hat G(t,\omega)
\mathcal{D}_{r'} \hat G(t,\omega)^\dagger \mathcal{H}(t)
\mathcal{D}_{r} \right],
\label{eq:def_transfer_function}
\end{equation}
while the diagonal elements $f_{r,r}(t,\omega)$ are determined by the following
expression for the sum over the first index:
\begin{equation}
\begin{split}
\bar f_{r'}(t,\omega) &= \sum_r f_{r,r'}(t,\omega) \\
&= \frac{1}{2\pi} \Tr\left[
\left(
\hat G^\dagger \frac{d\mathcal{H}}{dt} \hat G
- \frac{d}{dt}\left(G^\dagger \mathcal{H}\hat G\right)
\right)
\mathcal{D}_{r'} \right].
\end{split}
\label{eq:def_transfer_function_sum}
\end{equation}
Equation \eqref{eq:local_heat_freq_2} is the central result of this article.
It is a fully general expression for the local heat currents valid for arbitrary
temperatures and driving protocols. Although it has been derived based on Eq.
\eqref{eq:cov_matrix_fourier}, which is in principle only valid for
circuit descriptions that can be quantized, nothing prevents the evaluation
of Eqs. \eqref{eq:local_heat_freq_2}, \eqref{eq:def_transfer_function} and
\eqref{eq:def_transfer_function_sum} for general, overdamped circuits.
In fact, as we show analytically in Appendix \ref{ap:overdamped_static_heat},
the overdamped limit of the transfer function $f_{r, r'}(\omega)$ for a underdamped
circuit correctly matches the transfer function directly obtained from the corresponding
overdamped circuit. Later this is also verified numerically for the cooling scheme
of Section \ref{sec:example}.

To clarify the physical interpretation of the previous expressions we analyze
first the particular case of time independent circuits.

\begin{figure*}
  \includegraphics[width=\textwidth]{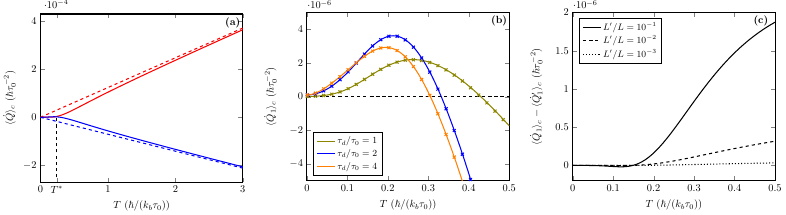}
  \caption{
  (a) Comparison of classical (dashed lines) and quantum (solid lines) heat currents for
  the circuit of Figure \ref{fig:example} ( $\Delta C/C = 10^{-1}$,
  $\omega_d = 10^{-2} (2 \pi/\tau_0)$, and $\tau_d = 2\tau_0$).
  (b) $\meann{\dot Q_1}_c$ as a function of $T$ for different values of $\tau_d/\tau_0$.
  Solid lines correspond to the heat currents computed for the overdamped circuit of Figure \ref{fig:example},
  while crosses correspond to the circuit of Figure \ref{fig:example_under} with $L'/L = 10^{-3}$
  (c) Difference between the heat currents obtained for the two circuits for different values of $L'/L$.}
  \label{fig:quantum_results}
\end{figure*}

\subsection{Undriven circuits}
If the matrix $\mathcal{H}$ is time independent, then for sufficiently long
times we have, from Eq. (\ref{eq:evol_G}), $\hat G(t,\omega) = \hat G_0(\omega) =
(i\omega - \mathcal{A}\mathcal{H})^{-1}$ (this is just the Laplace's transform
of the Green's function evaluated at $i\omega$). Then, asymptotically, the
transfer functions $f_{r,r'}(\omega)$ are time independent and $\bar f_{r'}(\omega) = 0$,
so that $f_{r,r}(\omega) = -\sum_{r'\neq r} f_{r',r}(\omega)$. Also,
in this case $f_{r,r'}(\omega)$ is symmetric under interchange of the indexes $r$ and $r'$.
To see this it is necessary to consider the block structure of the matrix $\mathcal{A}$,
that is inherited by the matrix $\hat G_0(\omega)$, and of the matrices
$\mathcal{D}_r$.
This is discussed in detail in Appendix \ref{ap:block_struct}, in connection with
the invariant nature of the dissipation upon time inversion.
Thus, using these properties we recover the usual Landauer-B\"{u}ttiker
expression for the heat currents \cite{rego1998,dhar2006,yamamoto2006}:
\begin{equation}
\meann{\dot{Q}_r} = \mean{j_r}\!\mean{v_r} + \!\sum_{r'}
\int_{-\Lambda}^{+\Lambda} \!\!\!\!\!\!\! d\omega \:\hbar \omega
\: f_{r,r'}(\omega)\left(N_{r'}(\omega) \!-\!  N_r(\omega)\right).
\label{eq:local_heat_undriven}
\end{equation}
From this equation, the quantity $f_{r,r'}(\omega) d\omega$ can be naturally
interpreted as the rate at which an excitation with frequency
between $\omega$ and $\omega + d\omega$ is transported from resistor $r'$ to $r$.
We note that the symmetry of the transfer function in the undriven case causes the
heat currents to only depend on the differences $(N_{r'}(\omega) +1/2) - (N_r(\omega)+1/2)$.
The 1/2 terms added to each Planck's distribution cancel each other.
However, this is not the case for driven circuits, where $f_{r, r'}(\omega)$
is not symmetric in general. In that case, the ground state fluctuations represented
by the $1/2$ term are responsible for the dissipation of heat into the resistors
due to the parametric driving even if all the temperatures are zero.

Thus, Eq. \eqref{eq:local_heat_freq_2} can be considered as the generalization to arbitrary
driving protocols of the Landauer-B\"{u}ttiker formula for the static case,  Eq. \eqref{eq:local_heat_undriven}.
In the following section, we provide simplified expressions for the transfer
functions in the case where the external driving is periodic. They are useful
for the numerical study of thermal cycles.

\subsection{Periodically driven circuits}
\label{sec:periodic_driving}

We now consider the situation where the matrix $\mathcal{H}(t)$
is a periodic function of time, and therefore can be decomposed as a Fourier series:
\begin{equation}
\mathcal{H}(t) = \sum_{k=-\infty}^{+\infty} \mathcal{H}_k \: e^{ik\omega_d t},
\label{eq:fourier_H}
\end{equation}
where $\omega_d$ is the angular frequency of the driving. In this case, assuming
stable dynamics and long times, the function $\hat G(t,\omega)$ is also periodic
with the same period of the driving, as shown in Appendix \ref{ap:periodic_driving}.
Thus, the following decomposition holds asymptotically
\begin{equation}
\hat G(t,\omega) = \sum_{j=-\infty}^{+\infty} \hat G_j(\omega)\: e^{ij\omega_d t}.
\label{eq:fourier_G}
\end{equation}
Then, the asymptotic covariance matrix of the system and the heat currents are also
periodic, with period $\tau = 2\pi/\omega_d$. We thus consider the average values
of the heat currents during a driving period, that we denote $\meann{\dot Q_r}_c$,
as we did in the example of Section \ref{sec:example}. They are given by
\begin{equation}
\meann{\dot Q_r}_c =  \mean{j_r}\! \mean{v_r} + \sum_{r'}
\int_{-\Lambda}^{+\Lambda} d\omega \: \hbar \omega \:
F_{r,r'}(\omega) \: (N_{r'}(\omega)+1/2),
\label{eq:local_heat_freq_periodic}
\end{equation}
where $F_{r,r'}(\omega)$ is the asymptotic average of $f_{r, r'}(t,\omega)$ during
a driving period, that can be expressed in terms of the Fourier components $\mathcal{H}_k$
and $\hat G_j(\omega)$:
\begin{equation}
F_{r,r'}(\omega) = \frac{1}{\pi} \sum_{j,j',k} \Tr\left[\mathcal{H}_k \hat G_j(\omega)
\mathcal{D}_{r'} \hat G_{j'}^\dagger(\omega) \mathcal{H}_{j'-j-k}
\mathcal{D}_{r}\right],
\label{eq:transfer_fourier_nondiag}
\end{equation}
for $r'\neq r$, and
\begin{equation}
\begin{split}
\bar F_{r'}(\omega) &= \sum_r F_{r,r'}(\omega) \\
&= \frac{1}{2\pi} \sum_{j, k} i k \omega_d \Tr\left[
\hat G_j^\dagger(\omega) \mathcal{H}_k \hat G_{j-k}(\omega)
\mathcal{D}_{r'} \right].
\end{split}
\label{eq:transfer_fourier_sum}
\end{equation}
Finally we note that, given the Fourier components $\mathcal{H}_k$ of the
external driving, the Fourier components $\hat G_j(\omega)$ of the Green's function
can be found by solving the following infinite set of algebraic equations
\begin{equation}
i(\omega +  j\omega_d) \hat G_j(\omega) = \mathds{1} \delta_{j,0} +
\mathcal{A} \sum_k \mathcal{H}_k \hat G_{j-k}(\omega),
\label{eq:algebraic_Gj}
\end{equation}
that is obtained by introducing the decompositions
of Eqs. \eqref{eq:fourier_H} and \eqref{eq:fourier_G} into Eq. \eqref{eq:evol_G}.
Some methods to solve this equation are discussed in Appendix \ref{ap:periodic_driving}.

The interpretation of the previous expressions in terms of elementary transport
processes is not as straightforward as in the regular Landauer-B\"{u}ttiker formula for
the undriven case. For open mechanical systems composed of quantum harmonic oscillators,
a physically clear decomposition of the local heat currents in terms of assisted transport and pair
creation of excitations was obtained recently \cite{freitas2017}. In particular,
the pair creation mechanism was shown to be dominant at low temperatures and to
be responsible for the ultimate limit for cooling in those systems.
In the next section, this quantum limit for cooling is illustrated numerically
for the cooling scheme introduced in Section \ref{sec:example}.

\subsection{Quantum limits for cooling}

\begin{figure}
\includegraphics[scale=.2]{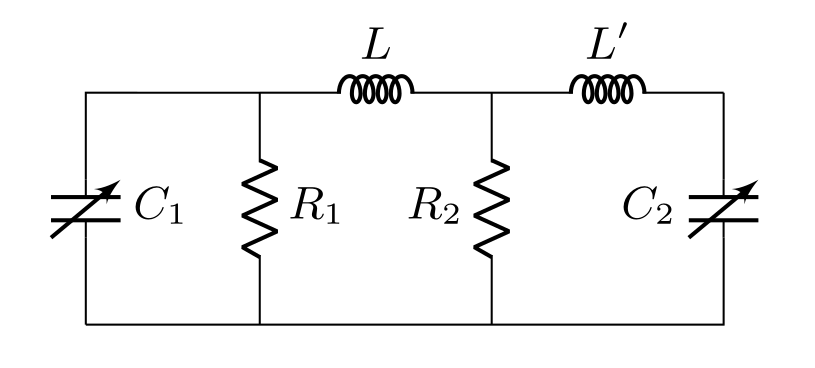}
\caption{A possible extension of the overdamped circuit of Figure \ref{fig:example}, where an additional
inductance $L'$ was introduced. In contrast to the circuit of Figure \ref{fig:example}, this circuit
can be canonically quantized.}
\label{fig:example_under}
\end{figure}

%

In this section we explore the low temperature behaviour of the heat currents
for the cooling scheme of Section \ref{sec:example}. We show how the quantum
corrections to the heat currents impose a minimum temperature below which
it is not possible to extract heat. Also, this will serve us to show that
the overdamped description of circuits can be directly employed to compute the heat
currents in the low temperature regime.

We consider the particular cooling scheme of Section \ref{sec:example} in
isothermal conditions. Thus, both resistors are at the same temperature $T$, and
we choose a driving amplitude $\Delta C$, frequency $\omega_d$, and phase
difference $\theta$ such that for high temperatures heat is extracted from resistor $R_1$
(thus, $\meann{\dot Q_1}_c < 0$), while it is dumped in resistor $R_2$ ($\meann{\dot Q_2}_c > 0$).
The heat currents are computed by evaluating Eqs. \eqref{eq:local_heat_freq_periodic},
\eqref{eq:transfer_fourier_nondiag} and \eqref{eq:transfer_fourier_sum}, based
on the overdamped circuit of Figure \ref{fig:example}.
In Figure \ref{fig:quantum_results}-(a) we compare the classical
(i.e, high temperature) heat currents, obtained by averaging Eq. (\ref{eq:local_heat}) over a driving
period, with the quantum heat currents according to Eq. (\ref{eq:local_heat_freq_periodic}).
The parameters are $\Delta C/C = 10^{-1}$, $\omega_d = 10^{-2} (2 \pi/\tau_0)$
and $\tau_d = 2\tau_0$ (recall that $\tau_0 = \sqrt{LC}$ and $\tau_d = RC$
are the oscillation and dissipation time scales, respectively). As expected,
the quantum heat currents approach the classical ones for increasing temperature.
However, below a given value of $T$ (indicated as $T^*$ in Fig. \ref{fig:quantum_results}-(a)),
$\meann{\dot Q_1}_c$ becomes positive and cooling stops. This is shown in more
detail in Figure \ref{fig:quantum_results}-(b), where it is also clear that the
value of $T^*$ decreases with increasing $\tau_d/\tau_0$, or equivalently with
decreasing dissipation rate (In this case $\omega_d$ is selected as the optimal driving
frequency for each value of $\tau_d/\tau_0$). This breakdown of the cooling effect is a strong coupling
result that cannot be captured with usual approaches based on master equations
(see \cite{karimi2016}, for example), as discussed in detail in \cite{freitas2017}.
All these results are independent of the cutoff frequency $\Lambda$ for large $\Lambda$.

In figure \ref{fig:quantum_results}-(b) we also show the results obtained based
on the circuit of Figure \ref{fig:example_under}, for $L'/L = 10^{-3}$ (indicated
by crosses). This circuit can be considered an extension of the one in Figure
\ref{fig:example} in which the additional stray inductance $L'$ was specified.
As shown in more detail in figure \ref{fig:quantum_results}-(c), the heat currents
obtained from the two descriptions match as $L'/L \to 0$. Thus, while overdamped
descriptions of circuits are not enough to construct a quantum model and compute the quantum state,
they are sufficient to compute the quantum corrections to the heat currents.

\section{Conclusions}

We presented a general study of the non-equilibrium thermodynamics
of driven electrical circuits. We derived the stochastic evolution of
the circuit state and of the heat currents dissipated in each resistor.
A relation between the topology of the circuit and the possibility of
defining finite heat currents under the white noise idealization was established.
As a first and simple example of application, we showed how to use our
formalism to study the transport and pumping of heat in a minimal circuit of two driven RC
circuits coupled by an inductor.

The initial classical treatment was then generalized in order to consider
the effects of quantum low-temperature noise. We considered a semiclassical
treatment in which the classical equations of motion are driven by noise
with a quantum spectrum. In contrast with treatments based on the full quantization
of the degrees of freedom in the circuit, our method has the advantage of being
directly applicable to circuits that cannot be quantized without the
additional specification of stray inductances or capacitances, but are
however detailed enough to properly describe the dynamics and also the thermodynamics.
Based on these results we expressed the heat currents for static circuits
in terms of the familiar Landauer-B\"{u}ttiker formula, and also obtained the generalization
of this expression for arbitrary driving protocols.

Our results offer a general formalism to study and design thermodynamical processes
in electrical systems from a first principles perspective and working in strongly
non-equilibrium conditions, and also away from the adiabatic and weak coupling regimes.
A direct application of the expressions provided  and of
the techniques illustrated in this article is the automatic optimization of thermal cycles
in complex and large electrical circuits. This is particularly straightforward
in the regime of high temperatures, where optimal cycles can be obtained that
could be later refined to take into account quantum effects.

Finally, nontrivial networks with stochastic linear dynamics are commonly used to describe
various kinds of complex systems including biological ones \cite{gnesotto2019, mura2018}.
In principle any such network can be emulated by a suitable RLC circuit.
This means that not only our results may apply to a very broad class of systems, but also
that experimental studies of those models could be carried out using RLC circuits.

\section{Acknowledgments}
We acknowledge funding from the European Research Council project NanoThermo (ERC-2015-CoG
Agreement No. 681456).

\bibliographystyle{unsrt}
\bibliography{references.bib}

\appendix
\input{supp_material_sections.tex}

\end{document}

%% file: supp_material_sections.tex
\section{Construction of the loop and cutset matrices}
\label{ap:description_circuits}

In this section we give a detailed explanation of how the loop and cutset matrices
are constructed for the example of Figure \ref{fig:circuit}-(a). The fist step is to
identify a normal tree of the circuit graph (one possible normal tree is indicated in blue
in Figure \ref{fig:circuit}-(b)). By definition, all the capacitors
and voltage sources should be part of the normal tree, while all inductors and current sources
should be out of it. Also, as any tree, it cannot contain loops, and all the nodes should be
connected. Thus, the edges $V$, $C_2$ and $C_1$
should be part of the normal tree, while the edges $L$ and $I$ should be out of it.
To identify the edges $R_1$, $R_2$, $R_3$, and $R_4$, we see that $R_1$ should
be part of the tree, since otherwise the node at which $L$ arrives would be disconnected.
$R_2$ should also be part of the tree, since otherwise the node from which $I$ departs
would remain disconnected. Thus, $R_1$ and $R_2$ are twigs.
Finally, $R_3$ and $R_4$ cannot be both part of the tree,
since in that case they would form a loop with $V$ and $C_2$. They also cannot be
both outside the tree, since in that case the central node from which $L$ departs
would be disconnected. Thus, one of them should be a twig and the other a link.
We are free to choose between the two possible options, and
in the example of Figure \ref{fig:circuit}-(b) we have choosen $R_4$ as a twig.

\begin{figure}
\includegraphics[scale=.18]{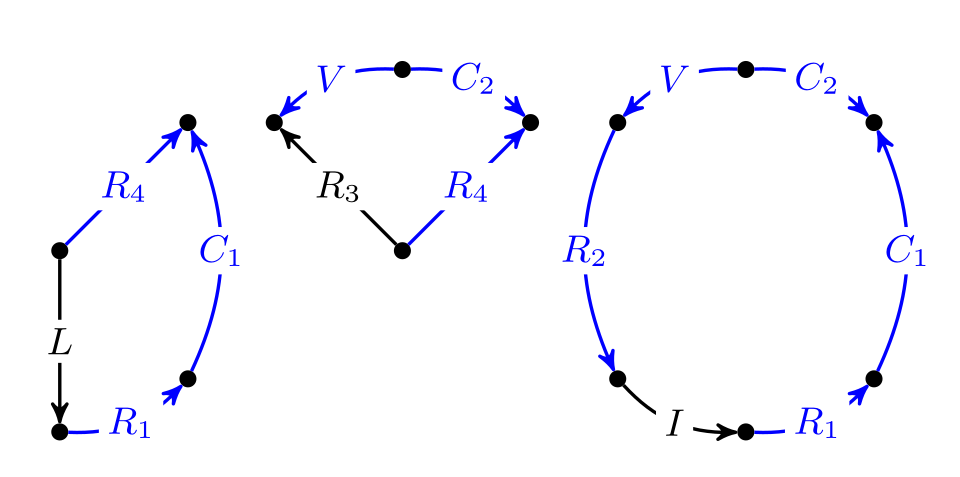}
\caption{Loops formed by adding a particular link to the normal tree.}
\label{fig:loop}
\end{figure}

Having identified the normal tree, we now proceed to construct the matrices of
fundamental loops and cutsets associated to it. We begin with the loops. We can
associate a loop to each link. For example, if we add link $L$ to the normal tree,
then a loop if formed by the edges $L$, $R_1$, $C_1$ and $R_4$ (See figure \ref{fig:loop}).
The orientation
of the loop is defined to be that of the added link, in this case $L$, which matches
the orientation of $R_1$ and $C_1$, but is contrary to that of $R_4$. The row
corresponding to this loop in the full loop matrix is then:
\begin{equation}
\begin{array}{c|ccccccccc}
& R_3 & L & I & \textcolor{blue}{V} & \textcolor{blue}{C_1} & \textcolor{blue}{C_2} & \textcolor{blue}{R_1} & \textcolor{blue}{R_2} & \textcolor{blue}{R_4}\\
 \hline
L &  0 & 1 & 0 & 0 & 1 & 0 & 1 & 0 & -1
\end{array},
\end{equation}
The rows corresponding to the links $R_3$ and $I$ are constructed analogously.
By adding link $I$ to the normal tree a loop is formed involving the
twigs $R_2$, $V$, $C_2$, $C_1$ and $R_1$. Adding link $R_3$ we form a loop
with twigs $V$, $C_2$ and $R_4$. All the loops are shown in Figure \ref{fig:loop}.
Taking into account their orientations, the final full loop matrix is:
\begin{equation}
B =
\begin{array}{c|ccccccccc}
& R_3 & L & I & \textcolor{blue}{V} & \textcolor{blue}{C_1} & \textcolor{blue}{C_2} & \textcolor{blue}{R_1} & \textcolor{blue}{R_2} & \textcolor{blue}{R_4}\\
\hline
R_3 &  1 & 0 & 0 & -1 & 0 & 1 & 0 & 0 & -1\\
L &  0 & 1 & 0 & 0 & 1 & 0 & 1 & 0 & -1\\
I &  0 & 0 & 1 & 1 & 1 & -1 & 1 & 1 & 0
\end{array}.
\end{equation}
We see that the first $3\times 3$ block is just the identity, and the rest of the
matrix is what is denoted by $B_\text{twig}$ in the main text.

\begin{figure}
\includegraphics[scale=.18]{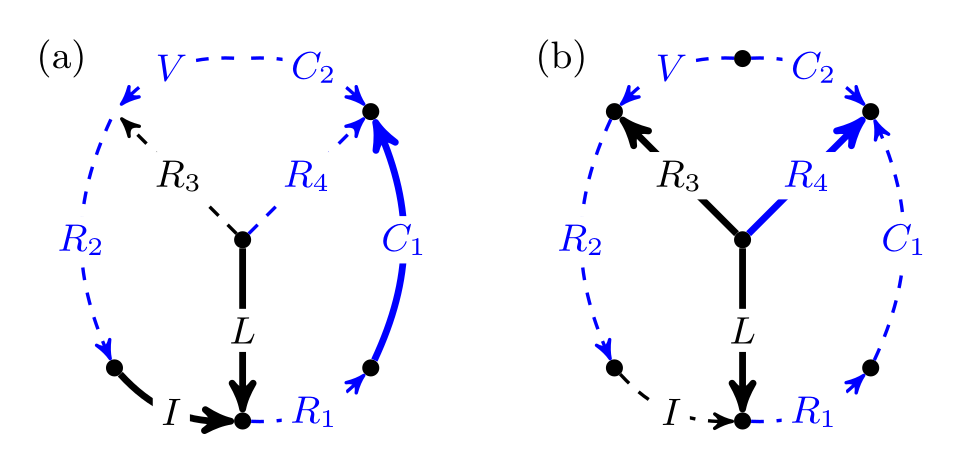}
\caption{
Cutsets corresponding to twig $C_1$ (a), and to twig $R_4$ (b).
The edges corresponding to a cutset are indicated with thick lines,
while the others are dashed.}
\label{fig:cutset}
\end{figure}

The construction of the cutset matrix is done as follows. A cutset can be associated
to each twig. For example, if we remove twig $C_1$, then the start and end nodes
of $R_1$ are disconnected from the others (in the tree, not in the full graph).
So we consider the splitting of the set of nodes in two subsets: the two nodes
connected to $R_1$, and all the rest. The edges going from one subset to the other
are $I$, $L$, and of course $C_1$ (see Figure \ref{fig:cutset}-(a)).
We give to this cutset the same orientation as $C_1$,
which happens to be contrary to the orientation of $I$ and $L$. Then, the
row corresponding to this particular cutset in the full cutset matrix is:
\begin{equation}
\begin{array}{c|ccccccccc}
& \textcolor{blue}{V} & \textcolor{blue}{C_1} & \textcolor{blue}{C_2} & \textcolor{blue}{R_1} & \textcolor{blue}{R_2} & \textcolor{blue}{R_4} & R_3 & L & I \\
\hline
\textcolor{blue}{C_1} & 0 & 1 & 0 & 0 & 0 & 0 & 0 & -1 & -1\\
\end{array}.
\end{equation}
The other rows corresponding to the twigs $V$, $C_2$, $R_1$, $R_2$ and $R_4$
are constructed in the same way. As an additional example, we note that removing
twig $R_4$ from the normal tree we isolate the central node, and the corresponding
cutset is formed by $R_4$, $R_3$ and $L$, which share the same orientation, as shown
in Figure \ref{fig:cutset}-(b).
Adding the corresponding row to the previous one we have:
\begin{equation}
\begin{array}{c|ccccccccc}
& \textcolor{blue}{V} & \textcolor{blue}{C_1} & \textcolor{blue}{C_2} & \textcolor{blue}{R_1} & \textcolor{blue}{R_2} & \textcolor{blue}{R_4} & R_3 & L & I \\
\hline
\textcolor{blue}{C_1} & 0 & 1 & 0 & 0 & 0 & 0 & 0 & -1 & -1\\
\textcolor{blue}{R_4} & 0 & 0 & 0 & 0 & 0 & 1 & 1 & 1 & 0\\
\end{array}.
\end{equation}
Filling in the remaining 4 rows to obtain the full cutset matrix, we find that
the first $6\times 6$ block is the identity, and the remaining block is the
one denoted by $Q_\text{link}$ in the main text.

\section{Necessary and sufficient condition to have well defined local heat currents for white noise
resistors}
\label{ap:topology_condition}

The matrix $\alpha^T \Pi_r R \alpha$ appearing in the last line of Eq. \eqref{eq:stoch_Qr_full}
has the following block structure:
\begin{equation}
\alpha^T \Pi_r R \alpha =
\left[
\begin{array}{cc}
A_r & B_r \\
-B_r^T & C_r
\end{array}
\right].
\end{equation}
The explicit form of each block can be derived from the definition of
$\alpha$ and using $2 \times 2$ block matrix inversion. Using this, we see that the
mean value of the last term in Eq. \eqref{eq:stoch_Qr_full} can be written as:
\begin{equation}
\begin{split}
&\mean{(- \Delta v_{R_l}^T \!+\! \Delta v_{R_t}^T Q_\text{RR}^T )
\: B_r \:
(Q_\text{RR} R_l^{-1} \Delta v_{R_l} \!+\! R_t^{-1} \Delta v_{R_t})} = \\
&2 k_b \delta(0) \left( \Tr\left[B_r Q_\text{RR} T_l\right]
+\Tr\left[Q_\text{RR}^T B_r T_t\right] \right),
\end{split}
\end{equation}
where $T_l$ and $T_r$ are diagonal matrices with the temperatures of the link
and twig resistors, respectively. For the previous expression to vanish for
arbitrary temperatures, the matrices $B_r Q_\text{RR}$ and $Q_\text{RR}^T B_r$
should have null diagonals.
The explicit form of $B_r$ is $B_r=(R_l + Q_\text{RR}^T R_t Q_\text{RR})^{-1}
Q_\text{RR}^T \pi_r (R_t^{-1} + Q_\text{RR} R_l^{-1} Q_\text{RR}^T )^{-1}$ if
the index $r$ correspond to a twig resistor, or
$B_r=-(R_l + Q_\text{RR}^T R_t Q_\text{RR})^{-1}
\pi_r Q_\text{RR}^T (R_t^{-1} + Q_\text{RR} R_l^{-1} Q_\text{RR}^T )^{-1}$
if $r$ correspond to a link resistor, where $\pi_r$ is the reduction of the projector
$\Pi_r$ to the appropriate twigs or links subspace. In any case, it is easy
to see that for the matrix $B_r Q_\text{RR}$ to have null diagonal
\emph{for any $r$ and for arbitrary values of the resistances}, all the components of the
matrix $Q_\text{RR}$ should vanish.

The condition $Q_\text{RR} =0$ means that there are no fundamental cutsets
associated to the normal tree containing simultaneously a link resistor and a
twig resistor. However, since it is ultimately related to the possibility of
defining finite heat currents, its validity should be independent of the choice of
normal tree. In fact, it is possible to obtain the following equivalent condition
that does not make any reference to a particular tree of the graph:
\begin{itemize}
  \item Given the full graph of the circuit, remove the I and L edges.
  \item Merge the connected components formed by E and C edges into a single node
  (this will remove internal edges, that might be resistors).
  \item The heat currents are finite if and only if the resulting network of only R edges is acyclic.
\end{itemize}
The equivalence with the condition $Q_\text{RR} = 0$ is given by the following two
observations: i) if there is a cycle in the final R network, then any normal
tree will have a fundamental cutset involving a resistor in the tree and a resistor
outside it, ii) if there is no cycle in the final R network, then all the edges in it
will be part of any normal tree, and the fundamental cutsets associated to them
will not involve any of the previously eliminated resistors (if any was eliminated).

\section{Block structure of the matrix $\mathcal{A}$ and the invariance of
dissipation upon time inversion}
\label{ap:block_struct}
In this section, we show explicitly how the block structure of the matrix $\mathcal{A}$
is related to the invariance of the dissipation upon time reversal and to
the symmetry property of the transfer function $f_{r,r'}(\omega)$ involved in the
expression for the heat currents for undriven circuits. We first recall
the necessary definitions:
\begin{equation}
\mathcal{A}(t) = \mathcal{M}_\text{cons} - \mathcal{M}_\text{diss}^T
{\alpha}(t)\mathcal{M}_\text{diss},
\label{apeq:def_A}
\end{equation}
\begin{equation}
\mathcal{M}_\text{cons} =
\left[
\begin{array}{cc}
& -Q_\text{CL}\\
Q_\text{CL}^T &
\end{array}
\right],
\qquad
\mathcal{M}_\text{diss} =
\left[
\begin{array}{cc}
-Q_\text{CR}^T&\\
& Q_\text{RL}
\end{array}
\right],
\end{equation}
and
\begin{equation}
{\alpha} =
\left[
\begin{array}{cc}
R_l & -Q_\text{RR}^T\\
Q_\text{RR} & R_t^{-1}
\end{array}
\right]^{-1}.
\end{equation}
Using $2 \times 2$ block matrix inversion, we see that $\alpha$ has also a block
structure like its inverse. Then it is straightforward to see that the
matrix $\mathcal{A}$ can be written as:
\begin{equation}
\mathcal{A} =
\left[
\begin{array}{cc}
\bm{s_1} & \bm{a}\\
-\bm{a}^T & \bm{s_2}
\end{array}
\right],
\label{eq:block_A}
\end{equation}
where $\bm{s_1}$ and $\bm{s_2}$ are symmetric matrices with dimensions $N_C\times N_C$
and $N_L \times N_L$, respectively ($N_C$ is the number of capacitors and $N_L$
the number of inductors). Therefore, the rate of energy dissipation is
\begin{equation}
\begin{split}
\dot{E}_\text{diss} &= \nabla E^T \mathcal{A} \nabla E =
x^T \mathcal{H} \mathcal{A} \mathcal{H} x  =\\
&= q^T C^{-1} \bm{s_1} C^{-1} q + \phi^T L^{-1} \bm{s_2} L^{-1} \phi,
\end{split}
\end{equation}
where in the second line we introduced the block structure of the
matrix $\mathcal{H} = \text{diag}(C^{-1}, L^{-1})$ and the state
vector $s=(q,\phi)^T$. Thus, we see that due to the block structure
of the symmetric part of $\mathcal{A}$, the rate of energy dissipation has
no cross terms coupling charges $q$ and fluxes $\phi$. Therefore it is
an even quantity under time reversal.

We now turn to discuss the symmetry of the transfer function $f_{r,r'}(\omega)$
for undriven circuits upon the interchange of $r$ and $r'$.
This function is given by (Eq. (\ref{eq:def_transfer_function}) in the main text):
\begin{equation}
f_{r,r'}(\omega) = \frac{1}{\pi} \Tr\left[\mathcal{H}\hat G(\omega)
\mathcal{D}_{r'} \hat G(\omega)^\dagger \mathcal{H}
\mathcal{D}_{r} \right],
\end{equation}
with $\hat G(\omega) = (i\omega \mathds{1} - \mathcal{A}\mathcal{H})^{-1}$.
We first note that since $\mathcal{H}$ is a positive definite matrix we can
write the product $\mathcal{H}\hat G(\omega)$ in the more symmetric form
$K(\omega) = \mathcal{H} \hat G(\omega) = \sqrt{\mathcal{H}}\left(i\omega \mathds{1} - \sqrt{\mathcal{H}}\mathcal{A}\sqrt{\mathcal{H}}\right)^{-1}\sqrt{\mathcal{H}}$, and thus we see that
if $\mathcal{A}$ were symmetric or antisymmetric, then the matrix $K(\omega)$ would inherit
that property. In any of those cases the function
\begin{equation}
f_{r,r'}(\omega) = \frac{1}{\pi} \Tr\left[K(\omega)
\mathcal{D}_{r'} K(\omega)^\dagger \mathcal{D}_{r} \right]
\end{equation}
would be trivially symmetric upon $r \leftrightarrow r'$. However, $\mathcal{A}$
has no definite symmetry. It has, nevertheless, a block structure that is also inherited
by $K(\omega)$:
\begin{equation}
K(\omega) = \mathcal{H} \hat G(\omega) =
\left[
\begin{array}{cc}
\bm{s'_1}(\omega) & \bm{a'}(\omega)\\
-\bm{a'}^T(\omega) & \bm{s'_2}(\omega)
\end{array}
\right],
\end{equation}
where the dimensions of $\bm{s'_1}(\omega)$, $\bm{s'_2}(\omega)$ and
$\bm{a'}(\omega)$ are the same as $\bm{s_1}$, $\bm{s_2}$ and  $\bm{a}$
in Eq. (\ref{eq:block_A}). If $K_s = \text{diag}(\bm{s'_1}, \bm{s'_2})$
and $K_a = K - K_s$ are the symmetric and antisymmetric part of $K$, we have
\begin{equation}
\begin{split}
f_{r',r} =&
\Tr\!\left[K_s \mathcal{D}_{r'} K_s^\dagger \mathcal{D}_{r}\right]+
\Tr\!\left[K_a \mathcal{D}_{r'} K_a^\dagger \mathcal{D}_{r}\right]+\\
&\Tr\!\left[K_s \mathcal{D}_{r'} K_a^\dagger \mathcal{D}_{r}\right]+
\Tr\!\left[K_a \mathcal{D}_{r'} K_s^\dagger \mathcal{D}_{r}\right],
\end{split}
\end{equation}
were we omitted the dependences in $\omega$. Finally from
the definitions of the matrices $\mathcal{C}_r$ it can be seen that
their products $\mathcal{D}_r = \mathcal{C}_r \mathcal{C}_r^T$ are block diagonal
whenever $Q_{RR} = 0$, which is the condition for the heat currents
to be properly defined in the white noise or high temperatures limit.
From this it follows that the two last traces in the previous equation are zero,
since their arguments have null diagonals. The remaining terms are
easily shown to be invariant upon $r \leftrightarrow r'$.

\section{Periodic driving}
\label{ap:periodic_driving}
In this section we show some useful properties of the Green's function of
periodically driven and stable circuits. We begin with the expression
for the transform $\hat G(t,\omega)$ given in Eq. \eqref{eq:def_G}:
\begin{equation}
\hat G(t,\omega) = \int_0^t  d\tau \: e^{-i\omega (t-\tau)} \: G(t,\tau).
\label{ap_eq:def_G}
\end{equation}
Here, $G(t,t')$ is the solution to
\begin{equation}
\frac{d}{dt} G(t,t') - \mathcal{A}(t)\mathcal{H}(t) G(t,t') = \mathds{1} \delta(t,t'),
\label{ap_eq:diff_G}
\end{equation}
with $G(t,t') = 0$ for $t<t'$. If $\mathcal{H}(t)$ is a periodic function with
period $\tau_d = 2\pi/\omega_d$, and the solution to the previous differential
equation is unique, then we have that $G(t,t') = G(t+\tau_d, t'+\tau_d)$.
Then:
\begin{equation}
\begin{split}
\hat G(t+\tau_d,\omega)
&= \int_0^{t+\tau_d} d\tau \: e^{-i\omega (t+\tau_d-\tau)} \: G(t+\tau_d,\tau)\\
&= \int_{-\tau_d}^{t} d\tau \: e^{-i\omega (t-\tau)} \: G(t+\tau_d,\tau+\tau_d)\\
&= \int_{-\tau_d}^{t} d\tau \: e^{-i\omega (t-\tau)} \: G(t,\tau)\\
&\simeq \int_{0}^{t} d\tau \: e^{-i\omega (t-\tau)} \: G(t,\tau) = \hat G(t,\omega),
\end{split}
\end{equation}
where in the first step we just employed a change of variables ($\tau \to \tau-\tau_d$),
and in the last step we assumed that the system is stable, in the sense that $G(t,t') \to 0$
for $|t-t'|\to \infty$. If that condition holds, then for sufficiently large $t$ we
can neglect the contribution of the first part of the integration domain. Thus,
we have shown that under this condition the function $G(t,\omega)$ is asymptotically
periodic, with period $\tau_d$.
We note that the stability condition does not always hold, even in the presence
of strong dissipation, since it is possible for the circuit to continuously absorb
energy from the driving and have a divergent dynamics. This is the phenomenon of
parametric resonance, that we exclude from our analysis.

Therefore, for long times $t$ we can give the following Fourier decomposition of the function
$\hat G(t,\omega)$:
\begin{equation}
\hat G(t,\omega) = \sum_{k=-\infty}^{k=+\infty} \hat G_j(\omega) e^{ij\omega_d t}.
\end{equation}
Then, inverting Eq. \eqref{ap_eq:def_G} and using Eq. \eqref{ap_eq:diff_G} (or equivalently,
transforming Eq. \eqref{ap_eq:diff_G} to obtain Eq. \eqref{eq:evol_G} in the main text),
we can find the following set of algebraic equations for the coefficients $\hat G_j(\omega)$:
\begin{equation}
i(\omega +  j\omega_d) \hat G_j(\omega) = \mathds{1} \delta_{j,0} +
\mathcal{A} \sum_{k=-\infty}^{k=+\infty} \mathcal{H}_k \hat G_{j-k}(\omega).
\label{ap_eq:algebraic_Gj}
\end{equation}
A simple method to solve these equations is to use a perturbative
approach in which the strength of the driving is considered small, i.e, we assume
$|\mathcal{H}_k| \ll |\mathcal{H}_0|$ for all $k\neq 0$. Then, to first order
in $\mathcal{H}_{k\neq 0}$ we have
$G_0(\omega) \simeq (i\omega\mathds{1} - \mathcal{A}\mathcal{H}_0)^{-1}$,
that is just the transform of the Green's function on the undriven circuit, and:
\begin{equation}
\hat G_j(\omega) \simeq \hat G_0(\omega+j\omega_d)\mathcal{A}\mathcal{H}_j \hat G_0(\omega)
\:\:\:\text{ for }\:\:\: j\neq 0.
\end{equation}
Higher orders in $\mathcal{H}_{k\neq 0}$ can be easily computed. From this solution,
we see that for weak driving the range of relevant Fourier components in $\hat G(t,\omega)$
is restricted by that of $\mathcal{H}(t)$. Thus, another non-perturbative method
to solve Eq. \eqref{ap_eq:algebraic_Gj} is just to truncate the Fourier space to some
maximum number of components given by $|k|\leq k_\text{max}$ and $|j| \leq j_\text{max}$
and then numerically solve the resulting finite system of linear equations for each value of $\omega$.

\section{Generalized Lyapunov Equation}
\label{ap:gen_lyapunov}

In this section we introduce a generalization of the Lyapunov equation that
is useful to compute the asymptotic state of periodically driven linear systems
subjected to white noise. For undriven circuits, the covariance matrix $\sigma$
for large times can be obtained as the solution of the following Lyapunov equation:
\begin{equation}
\mathcal{A}\mathcal{H} \sigma +
\sigma \mathcal{H}\mathcal{A}^T +
\sum_r 2k_bT_r \: \mathcal{C}_r\mathcal{C}_r^T = 0.
\label{ap_eq:static_lyapunov}
\end{equation}
For driven circuits there is no time-independent asymptotic state and one must
solve the dynamical equation:
\begin{equation}
\frac{d}{dt} \sigma(t) = \mathcal{A}\mathcal{H}(t) \sigma(t) +
\sigma(t) \mathcal{H}(t)\mathcal{A}^T + \sum_r 2k_bT_r \: \mathcal{C}_r\mathcal{C}_r^T.
\label{ap_eq:cov_matrix_evol}
\end{equation}
However, if the function $\mathcal{H}(t)$ is periodic, then we known from the results
of the previous section that the system state for large times will also be periodic
(with the same period of $\mathcal{H}$). Then we consider
$\mathcal{H}(t) = \sum_{k=-\infty}^{k=+\infty} \mathcal{H}_k e^{ik\omega_d t}$ and
introduce the following decomposition for $\sigma(t)$:
\begin{equation}
\sigma(t) = \sum_{k,k'=-\infty}^{+\infty} \sigma_{k,k'} \: e^{i(k-k')\omega_d t}.
\label{ap_eq:cov_matrix_fourier}
\end{equation}
Thus, the problem is now to find the coefficients $\sigma_{k,k'}$ in terms
of $\mathcal{H}_k$.
Note that, at variance with a regular Fourier decomposition, the previous expression
involves a double summation and as a consequence the coefficients $\sigma_{k,k'}$
are not uniquely defined. This choice, however, allows to cast
our problem as an extended Lyapunov equation. Indeed, it is useful to introduce
the following definitions:
\begin{equation}
A =
\left[
\begin{array}{ccccc}
\ddots & & & & \\
& \mathcal{AH}_0+i\omega_d\mathds{1}_n & \mathcal{AH}_{-1} & \mathcal{AH}_{-2} &\\
& \mathcal{AH}_1 & \mathcal{AH}_0 & \mathcal{AH}_{-1} &\\
& \mathcal{AH}_2 & \mathcal{AH}_1 & \mathcal{AH}_0-i\omega_d \mathds{1}_n &\\
& & & & \ddots \\
\end{array}
\right],
\end{equation}

\begin{equation}
S =
\left[
\begin{array}{ccccc}
\ddots & & & & \\
& \sigma_{-1,-1}^2 & \sigma_{-1,0}^2 & \sigma_{-1,1}^2 &\\
& \sigma_{0,-1}^2 & \sigma_{0,0}^2 & \sigma_{0,1}^2 &\\
& \sigma_{1,-1}^2 & \sigma_{1,0}^2 & \sigma_{1,1}^2 &\\
& & & & \ddots \\
\end{array}
\right],
\end{equation}
and
\begin{equation}
D_r =
\left[
\begin{array}{ccccc}
\ddots & & & & \\
& 0 & 0 & 0 &\\
& 0 & \mathcal{C}_r\mathcal{C}_r^T & 0 &\\
& 0 & 0 & 0 &\\
& & & & \ddots \\
\end{array}
\right].
\end{equation}
Then, it can be seen that the coefficients $\sigma_{k,k'}$ that give a solution
to Eq. \eqref{ap_eq:cov_matrix_evol} can be obtained by solving the following
generalized Lyapunov equation:
\begin{equation}
AS + SA^\dagger + \sum_r 2k_b T_r D_r = 0.
\label{ap_eq:generalized_lyapunov}
\end{equation}
Of course, to numerically solve this problem we need to truncate the dimensions of the
matrices $A$ and $S$. As we saw in the previous section, this is justified for
sufficiently weak driving. In the example given in Section \ref{sec:example}, the
function $\mathcal{H}$ has only $3$ Fourier components, and therefore to lower
order in $\mathcal{H}_{\pm 1}$ we can truncate the matrix $S$ to $3$ blocks in
each direction. After doing, this we solved symbolically the Lyapunov equation
of Eq. \ref{ap_eq:generalized_lyapunov} using Mathematica.

\section{Adiabatic and non-adiabatic decomposition of the entropy production}
\label{ap:ad_nonad_decomp}

The Fokker-Planck equation for the circuit state can be cast as
\begin{equation}
\frac{\partial}{\partial t} p(x,t)  = - \nabla^TJ(x,t) = - \nabla^T(J_c(x,t) + J_d(x,t)),
\label{apeq:FP_currents}
\end{equation}
where the total probability current $J(x,t)$ was split into conservative and
dissipative parts, that are respectively given by
\begin{equation}
J_c(x,t) = \mathcal{A}_a\mathcal{H}x \: p(x,t),
\end{equation}
and
\begin{equation}
J_d(x,t) = \mathcal{A}_s\mathcal{H}x \: p(x,t) - \sum_r 2k_b T_r \mathcal{D}_r \nabla p(x,t),
\end{equation}
where $\mathcal{A}_s$ and $\mathcal{A}_a$ are the symmetric and antisymmetric
parts of $\mathcal{A}$. The total entropy production rate is:
\begin{equation}
\dot \Sigma(t) = \sum_r \frac{1}{T_r} \int dx \: p(x,t)\:
j_r(x,t)^T \mathcal{D}_r j_r(x,t),
\label{apeq:sigma_total}
\end{equation}
with $j_r(x,t) = \mathcal{H}(t) x + k_b T_r \nabla \log(p(x,t))$. Using the FD
relation the dissipative probability current $J_d$ can be expressed in terms of the functions $j_r$
as follows:
\begin{equation}
\frac{J_d(x,t)}{p(x,t)}  = - \sum_r \mathcal{D}_r j_r(x).
\end{equation}

For a given instantaneous value of $\mathcal{H}(t)$, we define the corresponding
steady state distribution $p_\text{st}(x,t)$ as the one for which $\nabla^T J_\text{st}(x,t) = 0$.
It is the probability distribution to which the circuit would eventually relax if
the parameters are frozen at the values given by $\mathcal{H}(t)$. The functions $j_r(x,t)$
corresponding to $p_\text{st}(x,t)$ are denoted as $j_r^\text{st}(x,t)$.
Then, by replacing $j_r \to j_r - j_r^\text{st} + j_r^\text{st}$ in
Eq. \eqref{apeq:sigma_total}, we obtain the following decomposition of
the total entropy production:
\begin{equation}
\dot \Sigma = \dot \Sigma_\text{ad} + \dot \Sigma_\text{nad}  + \dot \Sigma^{'}_\text{nad},
\end{equation}
where
\begin{equation}
\dot \Sigma_\text{ad} = \sum_r \frac{1}{T_r} \int dx \: p(x,t)\:
j_r^\text{st}(x,t)^T \mathcal{D}_r j_r^\text{st}(x,t) \geq 0,
\end{equation}
\begin{equation}
\dot \Sigma_\text{nad} = \sum_r \frac{1}{T_r} \int dx \: p(x,t)\:
\Delta j_r(x,t)^T \mathcal{D}_r  \Delta j_r(x,t) \geq 0,
\end{equation}
with $\Delta j_r = j_r - j_r^\text{st}$, and finally
\begin{equation}
\dot \Sigma^{'}_\text{nad} = \sum_r \frac{1}{T_r} \int dx \: p(x,t)\:
\Delta j_r(x,t)^T \mathcal{D}_r j^\text{st}_r(x,t).
\end{equation}
The terms $\dot \Sigma_\text{nad}$ and $\dot \Sigma^{'}_\text{nad}$ vanish in
the adiabatic limit of infinitely slow driving, since the state is always the
stationary one and $\Delta j_r \to 0$. For the same reason they vanish for long
times if there is no driving. In contrast, $\dot \Sigma_\text{ad}$ converges to
the entropy production in the stationary state, which is $\sum_r \meann{\dot Q_r}/T_r$.
The non-adiabatic (nad) terms can be interpreted in terms of the relative
entropy of the instantaneous state with respect to the stationary state.
Indeed, $\dot \Sigma_\text{nad}$ can be expressed as:
\begin{equation}
\dot \Sigma_\text{nad} = -k_b \int dx \: \frac{\partial}{\partial t} p(x,t)\:
\log\left(\frac{p(x,t)}{p_\text{st}(x.t)}\right),
\end{equation}
and therefore in the absence of external driving ($\mathcal{H}(t)$ and therefore
$p_\text{st}(x,t)$ are constants) it equals $-k_b$ times the derivative of the relative
entropy $H(p|p_\text{st})$.
The term $\dot \Sigma^{'}_\text{nad}$ also accepts the following
expressions:
\begin{equation}
\begin{split}
\dot \Sigma^{'}_\text{nad} &= -k_b \int dx \: \frac{p(x,t)}{p_\text{st}(x,t)}\:
\nabla^T (p_\text{st}(x,t)) \mathcal{A}_a\mathcal{H}x\\
&= -k_b \int dx \: \nabla^T (p(x,t)) \mathcal{A}_a\mathcal{H} x \:
\log\left(\frac{p(x,t)}{p_\text{st}(x.t)}\right)\\
&= -k_b \int dx \: \nabla^T J_c(x,t) \:
\log\left(\frac{p(x,t)}{p_\text{st}(x.t)}\right),\\
\end{split}
\end{equation}
and can therefore be interpreted as a the change in the relative entropy $H(p|p_\text{st})$
due to the conservative flow $J_c$ in phase space.
At variance with $\dot \Sigma_\text{nad}$, $\dot \Sigma^{'}_\text{nad}$ is not
always positive. It vanishes identically in the following cases: i) for circuits
with no inductors or no capacitors (in that case we can consider $\mathcal{A}_a =0$
and the dynamics is always overdamped), and ii) in isothermal conditions,
since the steady state satisfies $\nabla p_\text{st}/p_\text{st} \propto Hx$
and therefore $\dot \Sigma^{'}_\text{nad} \propto \Tr\left[\mathcal{A}_a \mathcal{H} \sigma \mathcal{H}\right] = 0$.
Finally, adding the last two equations and using Eq. \eqref{apeq:FP_currents},
we can see that the sum of the non-adiabatic
terms only depends on the dissipative current:
\begin{equation}
\dot \Sigma_\text{nad} + \dot \Sigma^{'}_\text{nad}  = k_b \int dx \: \nabla^TJ_d(x,t)
\: \log\left(\frac{p(x,t)}{p_\text{st}(x.t)}\right).
\end{equation}
We note that this quantity is not necesarily positive definite. It is positive
for overdamped circuits, since we have seen that in that case $\dot \Sigma_\text{nad} = 0$
and $\Sigma_\text{nad} \geq 0$.

\section{Underdamped limit for static heat conduction}
\label{ap:overdamped_static_heat}

\begin{figure}
\includegraphics[scale=.18]{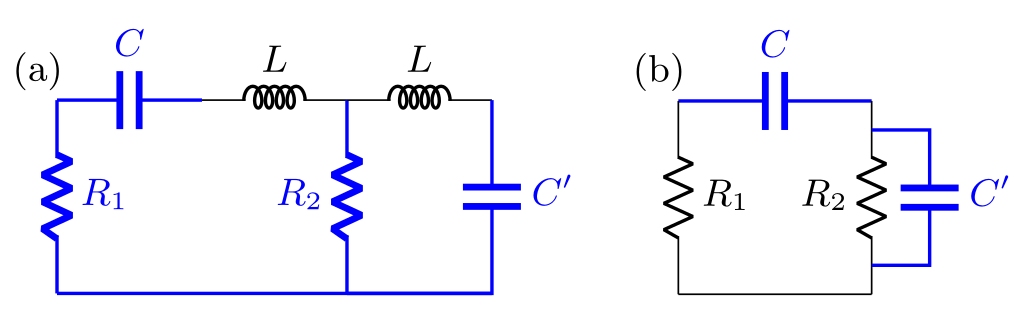}
\caption{(a) A underdamped circuit description that can be directly quantized. (b)
overdamped circuit obtained in the limit $L/R^2 C\to0$. Normal trees in each case are indicated in blue.}
\label{fig:overdamped_limit}
\end{figure}

In this section we illustrate with an example how the heat transfer function of
an overdamped circuit is recovered by taking the appropiate limit in the
heat transfer function of a underdamped circuit. We consider the circuit in
Figure \ref{fig:overdamped_limit}-(a), that in the limit $L/R^2 C\to0$ reduces to
the circuit of Figure 2-(c) in the main text.

We consider the heat transfer function given by Eq. \eqref{eq:def_transfer_function}
in the main text (only the non-diagonal element is necessary in this case of
stationary heat conduction):
\begin{equation}
f_{1,2}(\omega) = \frac{1}{\pi} \Tr\left[\mathcal{H}\hat G(\omega)
\mathcal{D}_{1} \hat G(\omega)^\dagger \mathcal{H} \mathcal{D}_{2} \right],
\end{equation}
where
\begin{equation}
\hat G(\omega) = (i\omega \mathds{1} - \mathcal{A}\mathcal{H})^{-1}.
\end{equation}
The matrices $\mathcal{A}$, $\mathcal{H}$ and $\mathcal{D}_{1/2}$ for each circuit
are to be constructed according to the procedure of Section II in the main text.
In this way, we obtain the following transfer function for the circuit of
Figure \ref{fig:overdamped_limit}-(a):
\begin{equation}
f^\text{(a)}_{1,2}(\omega) \!= \!
\frac{R_1 R_2}{\pi} \! \! \left| \frac{ i\omega L + (i\omega C')^{-1}}
{\left(i\omega L \!+\! \frac{1}{i\omega C} \!+\! R_1 \!+\! R_2\right)
\left(\frac{1}{i\omega C'} \!+\! R_2 \right) \!-\! R_2}\right|^2,
\end{equation}
and for the circuit of Figure \ref{fig:overdamped_limit}-(b):
\begin{equation}
f^\text{(b)}_{1,2}(\omega) = \frac{1}{\pi R_1 R_2} \left| \frac{i\omega C}
{\left(i\omega C \!+\! \frac{1}{R_1}\right)
\left(i\omega C' \!+\! \frac{1}{R_1} \!+\! \frac{1}{R_2}\right) \!-\! \frac{1}{R_1^2}}
\right|^2.
\end{equation}
Thus, a simple calculation shows that in fact
\begin{equation}
f_{1,2}^\text{(a)} (\omega) |_{L=0} = f_{1,2}^\text{(b)} (\omega).
\end{equation}